\documentclass{article}
\usepackage[utf8]{inputenc}

\usepackage{fullpage}
\usepackage{url}
\usepackage{multirow}
\usepackage{pdflscape}
\usepackage{paralist}
\usepackage[shortlabels]{enumitem}
\usepackage{tikz}
\usepackage{xspace}
\usepackage{wrapfig}
\usepackage{ulem}
\usepackage{svg}
\usepackage{tabularx, booktabs}
\usepackage{color, colortbl}
\usepackage{setspace}
\usepackage{xcolor}
\usepackage{makecell}
\usepackage{graphicx}
\usepackage{caption}
\usepackage{subcaption}

\newcommand{\revised}[1]{{\color[rgb]{0,0,0}#1}}

\def\steeringTitle{Dynamic Orchestration}
\def\steering{Dynamic Orchestration\xspace}
\def\multistageTitle{Multistage pipeline}
\def\multistage{Multistage Pipeline\xspace}
\def\inverseTitle{Inverse design}
\def\inverse{Inverse Design\xspace}
\def\replicaTitle{Digital replica}
\def\replica{Digital Replica\xspace}
\def\distributedTitle{Distributed models}
\def\distributed{Distributed Models\xspace}
\def\adaptiveTitle{Adaptive execution for training large AI models}
\def\adaptive{Adaptive Training\xspace}

\begin{document}

\title{AI-coupled HPC Workflow Applications, Middleware and Performance}

\author{Wesley Brewer, Ana Gainaru, Fr\'ed\'eric Suter, Feiyi Wang \\
  Oak Ridge National Laboratory, Oak Ridge, TN, USA
\and
Murali Emani \\ 
Argonne National Laboratory, Lemont, IL, USA
\and 
Shantenu Jha \\
Rutgers University, New Brunswick, NJ, USA\\
Princeton Plasma Physics Laboratory \& Princeton University, Princeton, NJ, USA
}
\date{}
\maketitle

\begin{abstract}
AI integration is revolutionizing the landscape of HPC simulations, enhancing the importance, use, and performance of AI-\revised{coupled} HPC workflows.
This paper surveys the diverse and rapidly evolving field of AI-driven HPC and provides
a common conceptual basis for understanding \revised{such workflows}. Specifically,
we use insights from different modes of coupling AI into HPC workflows to
propose six execution motifs most commonly found in
scientific applications. By definition, the proposed set of execution motifs is
incomplete and evolving. However, they allow us to analyze the primary performance challenges  \revised{across AI-integrated simulation environments}. We close with a listing of open challenges, research issues, and suggested areas of investigation, including the the need for specific
benchmarks that will help evaluate and improve the execution of AI-\revised{coupled} HPC
workflows.

\end{abstract}

\section{Introduction}

Multiple recent publications have demonstrated how artificial intelligence (AI) coupled with traditional
high performance computing (HPC) simulations can provide practical performance enhancements of
10\textsuperscript{3} or more~\cite{9103969, brace2022coupling, Clyde2023}. AI-coupled HPC workflows -- 
as opposed to decoupled AI and HPC -- 
involve the concurrent, real-time coupled execution
of AI and HPC tasks in ways that allow the AI systems to steer or inform the HPC
tasks and vice versa. 
The real-time coupling and concurrency of HPC and AI
tasks are fundamental, as they allow bidirectional influence.
The online coupling of an AI system to HPC workflows can be used for a variety
of scenarios, e.g., an AI-surrogate model to substitute part of
expensive simulation, an AI system to learn the function or determine
the parameters or the practical fields, and an AI system to guide  a campaign using an objective function or performance Pareto-optimal or
resource-optimal experiments (using, for example, Active/Reinforcement Learning).

Driving the need to couple AI systems to traditional HPC simulations is the
promise of enhancing the ``effective performance'' of HPC workflows, where
enhancement is measured by ``science for a given amount of computing''.
\revised{Such coupled} workflows continue to overcome the limitations of traditional
forward simulations in increasingly sophisticated and pervasive ways. Further,
\revised{these hybrid} workflows will overcome traditional bottlenecks that prevent
greater scale -- physical and temporal or higher resolutions. The integration
of AI into a computational workflow is also a sustainable and scalable way to
obtain significant performance gains and present an opportunity to avoid
simulation enhancements that are overly sensitive to processor
architecture~\cite{fox2019learning,jha2019understanding}. Put together,
AI-coupled HPC workflows present a promising paradigm~\cite{jha2022ai,scaling2023merchant,lam2023learning} which leverages the
ubiquitous interest and wide capabilities being developed for AI but employs
them to overcome performance bottlenecks due to unsustainable approaches.

Integrating AI systems into traditional high-performance computing workflows has demonstrably enabled highly accurate modeling and holds significant promise for accelerating scientific discovery. There exist multiple promising examples of workflows that couple AI methods
with traditional HPC workflows, typically by extending HPC workflows with
additional capabilities to support the concurrent execution of AI modules
~\cite{wang2023scientific,lavin2021simulation,covidisairborne2021ijhpca,wang2020machine}.
Multiple solutions have successfully coupled AI systems to be either ``about''
the HPC workflows or ``outside'' (typically in the outer loop or even possibly
remote from) the main HPC workflow. \revised{The integration of AI with HPC can be systematically understood through a taxonomy that defines three key modes of interaction: AI-in-HPC, AI-out-HPC, and AI-about-HPC~\cite{jha2022ai}. AI-in-HPC signifies instances where an AI model directly replaces or acts as a surrogate for an entire HPC simulation or one of its components, thereby accelerating computations. In contrast, AI-out-HPC describes scenarios where an AI model operates externally to the core HPC simulation loop but dynamically controls or orchestrates the progression of the HPC workflow, such as through active learning or reinforcement learning to optimize computational campaigns. Finally, AI-about-HPC refers to situations where AI models function concurrently and are coupled with the primary HPC tasks, serving to complement traditional computational methods by enhancing scientific results or improving overall efficiency without necessarily taking over core simulation functions.}

\begin{figure}[t!]
    \centering
    \includegraphics[width=0.8\textwidth]{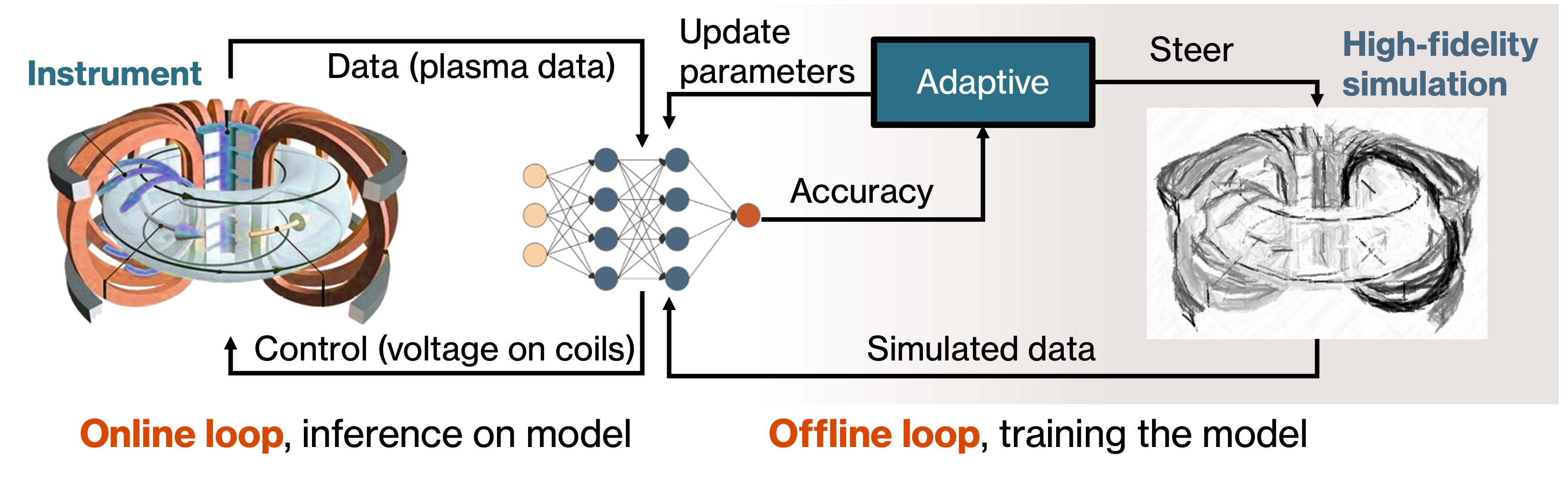}
    \caption{Example of an AI-coupled HPC workflow using models to replace expensive high-fidelity simulations for steering instruments}
    \label{fig:science-examples}
    \vspace{-\baselineskip}
\end{figure}

Figure~\ref{fig:science-examples} shows a typical science application where AI has advanced simulations and experiments by coupling AI to HPC
workflows via different mechanisms. 

\revised{Hybrid AI-HPC applications offer significant improvements over traditional high-performance computing (HPC) methods by fundamentally reshaping parallel computation, resource utilization, and scientific insight extraction. Unlike vanilla HPC applications, which rely heavily on exhaustive, synchronous computations across HPC platforms, AI-integrated workflows employ predictive inference, intelligent execution, and optimization to achieve scientific outcomes.

Traditional HPC methods necessitate step-by-step simulation runs, which can be expensive to explore even modest design spaces. In contrast, AI-guided simulations leverage trained AI models to predict system behaviors, significantly reducing computational burdens. AI models, after training on extensive datasets from high-fidelity simulations, can almost instantaneously forecast outcomes, bypassing the exhaustive computations previously required. While the initial AI model training might demand significant computational resources, subsequent inferences are orders of magnitude faster and more cost-effective. This enables real-time optimizations, rapid design-space exploration, and immediate predictions, thereby dramatically streamlining workflows and accelerating scientific discoveries.

Furthermore, AI frameworks naturally employ asynchronous computation, fundamentally differing from traditional HPC's  synchronous, lockstep patterns. AI workloads, which can tolerate and even thrive on asynchronous updates, such as certain neural network training algorithms, minimize synchronization overheads, a critical bottleneck in large-scale HPC scenarios. This asynchronous execution allows different parts of computations to progress independently, significantly improving efficiency by not tying the entire system to the slowest processing elements.

Integrating AI into HPC workflows restructures parallel algorithm designs and communication patterns, directly addressing and overcoming the strong scaling limits inherent to traditional HPC simulations. AI intelligently predicts outcomes, identifies critical computation regions, and streamlines workflows by learning underlying physical laws directly from data or encoding known physics into AI models. These predictive and adaptive capabilities minimize communication overhead, balance computational loads, and accelerate computationally intensive segments without sacrificing accuracy, thereby pushing beyond traditional scaling limitations and significantly enhancing scalability.

AI-coupled HPC workflows support algorithms/methods where machine learning complements traditional solvers by initializing simulations, preconditioning algorithms, or estimating error bounds, enhancing both computational efficiency and scientific reliability. Building on this integration, surrogate modeling allows AI to approximate complex simulations,  delivering results significantly faster than conventional approaches. Through accelerated simulations, AI focuses computational resources on the most promising regions of parameter space using techniques like active learning, thereby reducing cost and time. Hybrid approaches as exemplified above, yield faster scientific insights, broader design explorations, representing an essential shift away from traditional performance gains and limitations of strong scaling.}

The focus in this paper is on the scientific workflows that involve the online (real-time)
coupling and concurrent execution of AI and HPC, not offline training
and inference. In addition to providing an overview of existing AI-coupled HPC workflows, the
primary objectives of this paper are to identify classes of AI-coupled HPC
workflows and to provide an overview of the software systems used to support
them. The paper also discusses performance and systems challenges, identifying
robust solutions and lacking ones. Finally, the paper highlights some
open research issues to advance AI-HPC workflows.

The paper is organized as follows: Section~\ref{sec:motifs}
reviews coupling modes between AI and HPC tasks and introduces six execution
motifs that describe the interaction and coordination patterns of AI and HPC
components. These motifs are used to understand a range of AI-\revised{coupled} HPC
workflow applications and \revised{their associated} performance challenges. Execution motifs are important at multiple levels,
not the least of which is their influence on the design of application-level
software systems and task runtime systems. For instance, the frequency of
interaction in a given motif informs the performance requirements and the
optimizations available to lower-level libraries. Further, the diversity of AI
approaches and model types mean intrinsic variation in training and execution
times. Section~\ref{sec:frameworks} presents several frameworks and libraries
developed to ease the coupling of AI and HPC components in a single workflow
and illustrates how each of these frameworks relates to the proposed
high-level modes of execution motifs. Section~\ref{sec:performance} discusses
various performance issues related to AI-\revised{coupled} HPC workflows -- organized around the
motifs -- such as load balancing, workload management, and dataflow
performance bottlenecks. We conclude in Section~\ref{sec:open} with a
discussion of several open issues and challenges that \revised{this emerging class of} workflows present\revised{s}.

\section{Execution Motifs}\label{sec:motifs}

The concept of AI-coupled HPC workflows is well-established, with a current taxonomy dividing them into three primary modes: AI-in-HPC, AI-out-HPC, and AI-about-HPC~\cite{jha2022ai,jha2019understanding}.
These modes are characterized by the relative placement of AI methods concerning HPC simulations.
\textbf{AI-in-HPC} represents the scenario in which an AI system is introduced instead of a component of the HPC simulation, or possibly, instead of the whole simulation
itself, i.e., the AI model serves as a ``total
surrogate''~\cite{Kasim20published}. \textbf{AI-out-HPC} captures situations
where an AI system resides ``outside'' of the traditional HPC simulation loop
but dynamically controls the progression of the HPC workflow~\cite{tao2021nanoparticle,yao2021inverse} (e.g., control of computational campaigns via reinforcement learning). Finally,
\textbf{AI-about-HPC} represents the situation where AI systems are concurrent
and coupled to the main HPC tasks: analysis and training codes use the output of the HPC simulation to provide further insights. 
This section expands on this taxonomy. 
Given the increasing trend and benefits of combining these non-mutually exclusive modes~\cite{fox2019learning,fox2019taxonomy,jha2019understanding}, the need for compatible software solutions for managing their execution and data exchange is paramount.
For this purpose, we build on the concept of coupling modes (AI-x-HPC, where x could be in, out,
or about), and we introduce execution motifs. Whereas modes provide insight into
the static coupling of AI systems with HPC simulations, execution motifs
capture and provide information on recurring dynamic interaction patterns between the AI systems and the HPC simulations, driven by a high-level functional goal.

We define an execution motif using the characteristics the following characteristics: 
\begin{compactitem}
    \item The first class of characteristics defines the \textbf{interaction patterns} between the AI and HPC components: (1) data-flow type (e.g., one-to-one or one-to-many) and direction (i.e., HPC-to-AI or AI-to-HPC or both); (2) control flow type and direction (e.g., one/many AI components steering one or many HPC simulations, HPC simulation guiding the AI training process, etc.); and (3) user-involvement in the process.
    \item The second class defines the \textbf{coupling patterns} between AI and HPC components, namely: (4) concurrency requirements between the AI and HPC components (e.g., real-time interaction or post-mortem analysis); (5) dynamism requirements in the composition of the workflow (e.g., spawning/terminating new processes or training sequences) and in the interaction patterns (e.g., data/control flow keeps changing, new connections are spawned/terminated); and (6) federation requirements (e.g, network requirements between AI and HPC components or  heterogeneity in resources).
    \item Each motif is associated with a \textbf{scope} which represents its primary purpose (e.g., is AI used to improve or optimize HPC components, is HPC used to build improved AI models, or both)
\end{compactitem}

\begin{figure}[t!]
    \centering
    \includegraphics[width=0.9\textwidth]{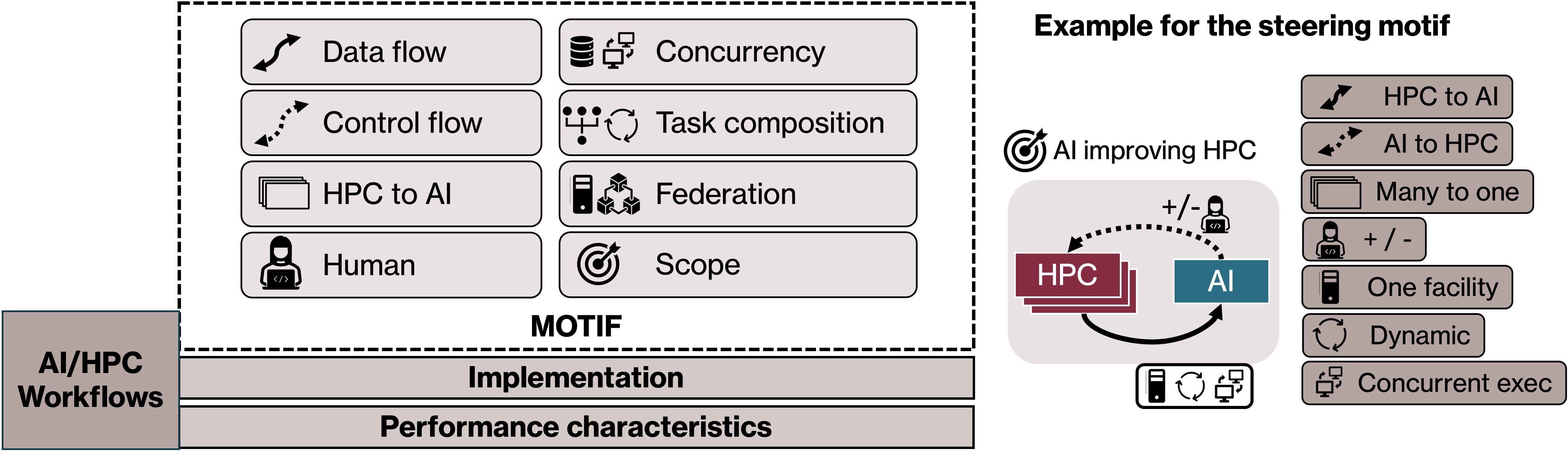}
    \caption{Defining characteristics of AI-coupled HPC workflows: (left) Breakdown of the main characteristics used to define the different behavioral motifs; (right) Illustrative example of the characteristics typically used in the steering motif following the breakdown presented on the left.}
    \label{fig:overview}
    \vspace{-\baselineskip}
\end{figure}

These execution motifs serve as identifiers for (groups of) AI-coupled HPC workflows that exhibit
similar interaction and coupling patterns between AI and HPC components, and
thus encounter similar performance bottlenecks and require similar optimizations when
deployed on HPC systems. 
\revised{Because a workflow might incorporate aspects of several motifs, grasping the relationships between workflow patterns, the frameworks employed, and the performance considerations outlined in this paper empowers users to better comprehend the available options and choose appropriate solutions tailored to their multi-faceted workflow needs.}
\revised{Figure~\ref{fig:overview} illustrates the characteristics used to differentiate between different workflow patterns and how they can be combined to create an example motif. }

Based on a literature survey of large-scale workflows, we identify six recurring execution motifs that capture the dominant coupling patterns between AI and HPC. These motifs reflect the dynamic interaction styles and purposes of AI within scientific workflows. Table~\ref{table:motifs} summarizes the distinct patterns of each motif according to the characteristics introduced in Figure~\ref{fig:overview}. We briefly list them here:

\begin{enumerate}
\item \textbf{Dynamic Orchestration:} AI steers ensembles of HPC simulations, dynamically spawning or terminating runs.
\item \textbf{Multistage Pipeline:} AI-based filters guide execution across stages in a pipelined HPC campaign.
\item \textbf{Inverse Design:} AI iteratively infers simulation inputs from observations to solve inverse problems.
\item \textbf{Digital Replica:} AI digital replicas interact with simulations for prediction, monitoring, and feedback.
\item \textbf{Distributed Models:} AI and HPC components operate across geographically distributed resources.
\item \textbf{Adaptive Training:} HPC simulations generate data that adaptively improve AI models during training.
\end{enumerate}

\begin{figure}[t!]
    \centering
    \includegraphics[width=0.8\textwidth]{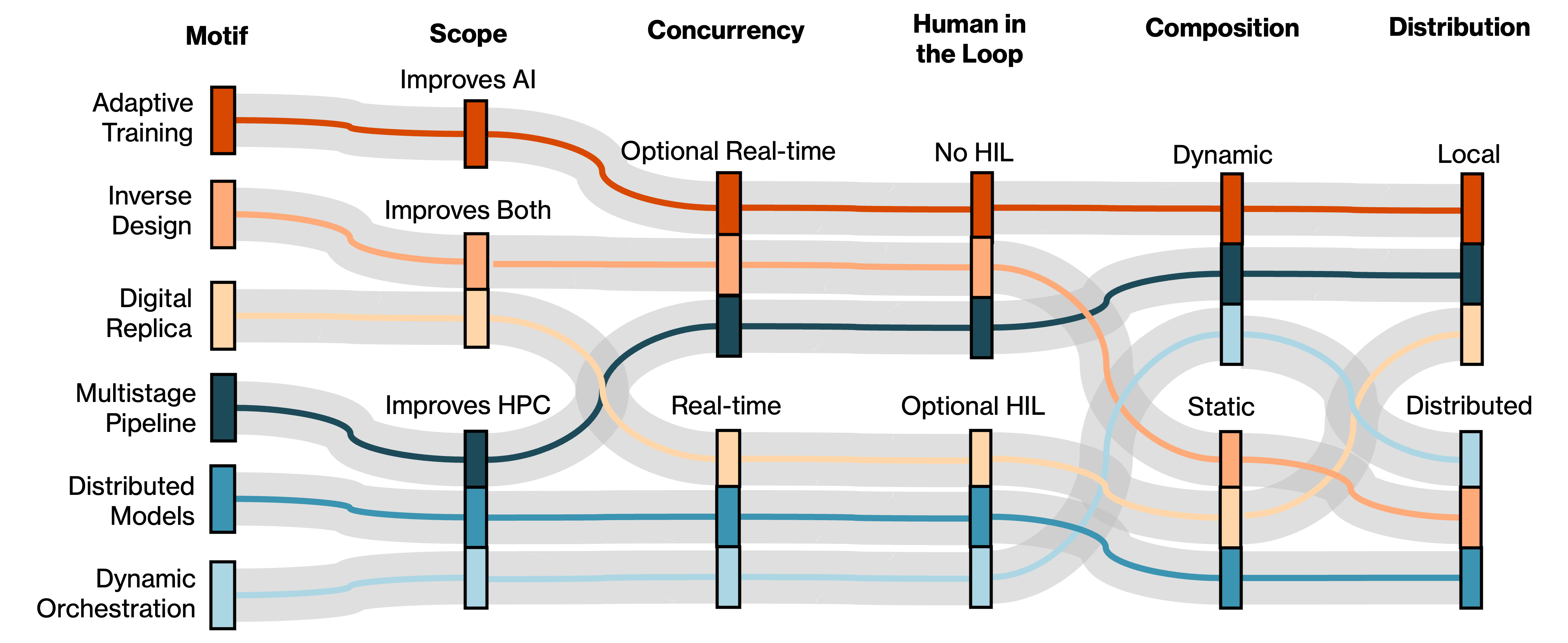}
    \caption{Sankey diagram of the motifs and their characteristics.}
    \label{fig:sankey}
    \vspace{-\baselineskip}
\end{figure}

\begin{table}[ht!]
\centering
\resizebox{\textwidth}{!}{
\begin{tabular}{|p{2.1cm}|p{7.5cm}|p{5.8cm}|}
\hline
\textbf{Motif} & \textbf{Patterns} & 
\hfill\textbf{Example Use Case} \hfill\,\hfill 
\\\hline  \hline 

\textbf{Dynamic\newline Orchestration} & \raisebox{-\totalheight}{\includegraphics[width=7.5cm]{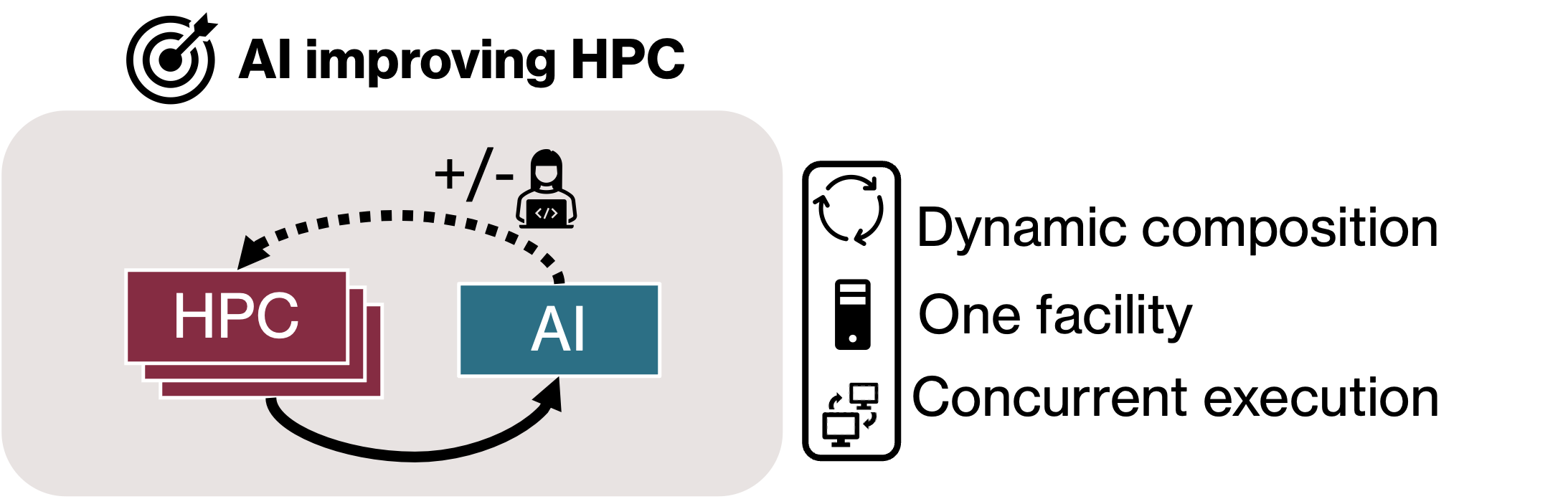}}
                   &  
                   - Command-and-control of physical experiments and simulations (e.g. between shots feedback for plasma physics) 
                   \\\hline

\textbf{\multistage} & \raisebox{-\totalheight}{\includegraphics[width=7.5cm]{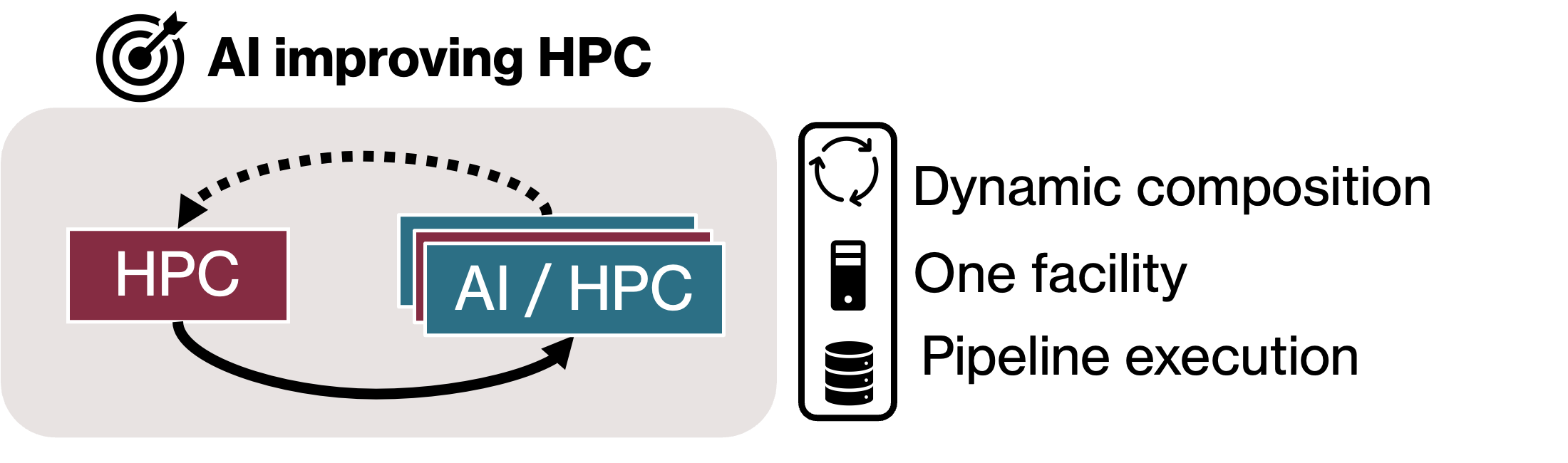}} 
                   & 
                   - Large-scale MD simulations using AI sampling of a system with many degrees of freedom\\ \hline
\textbf{Inverse\linebreak Design} & \raisebox{-\totalheight}{\includegraphics[width=7.5cm]{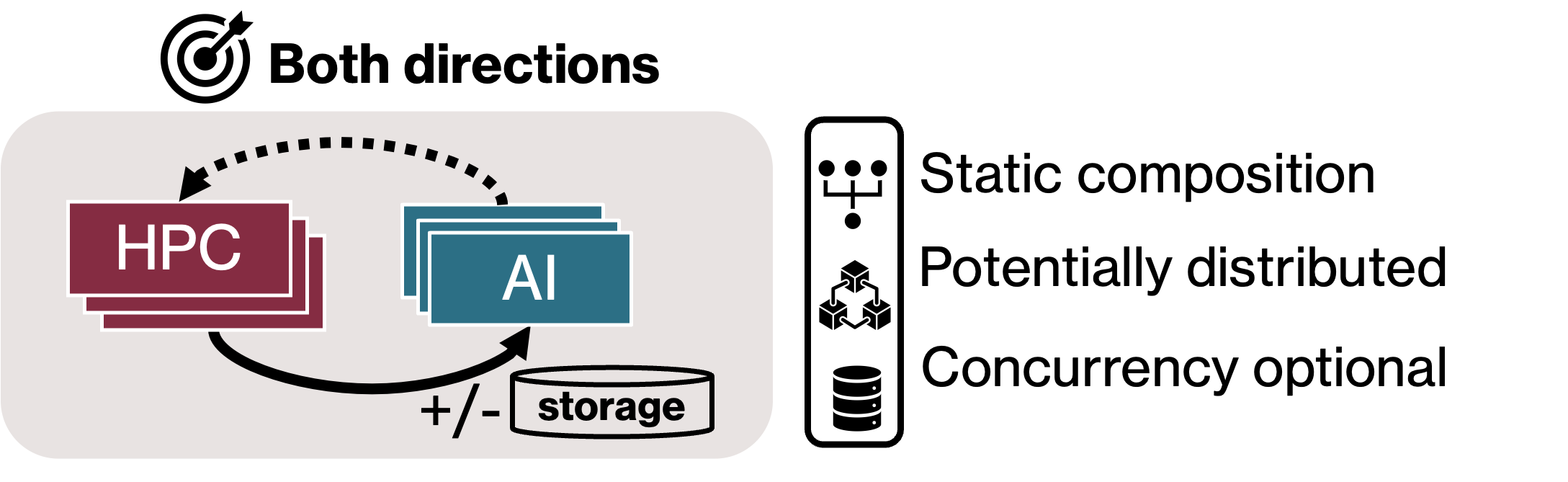}} 
                   & 
                   - Materials discovery to address the problem of data sparsity and reduce the need for domain-specific knowledge \\ \hline
\textbf{\replica} & \raisebox{-\totalheight}{\includegraphics[width=7.5cm]{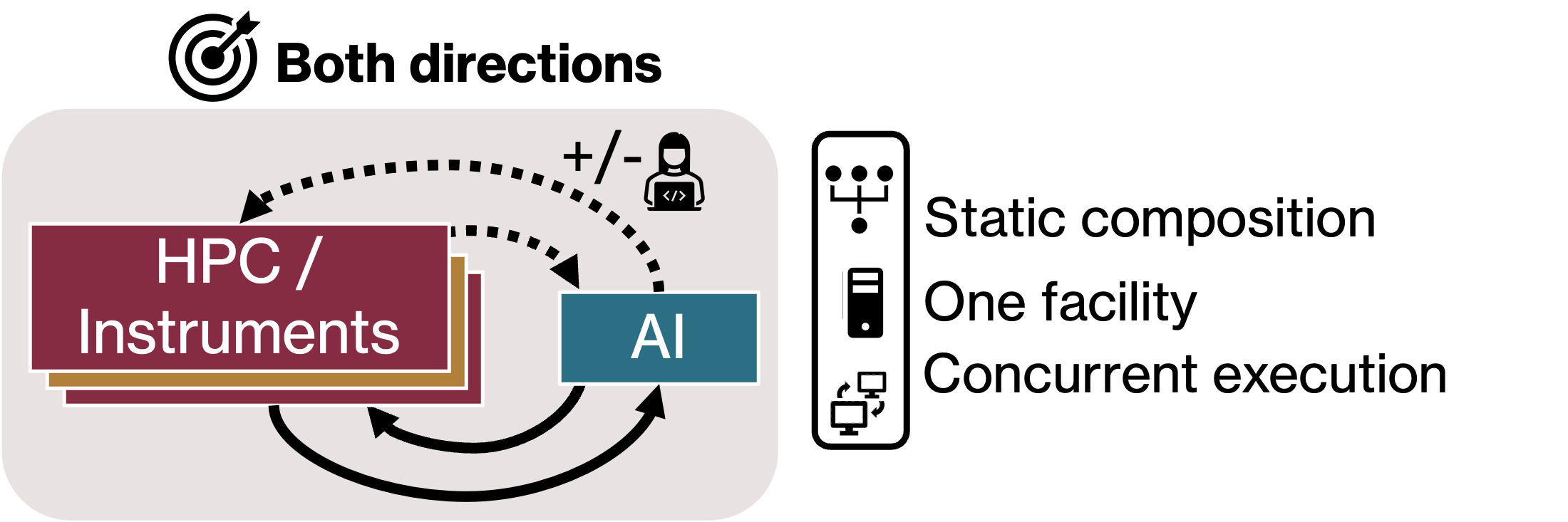}}
                   & 
                   - Digital twin of a fusion reactor running concurrent with ITER~\cite{HOLTKAMP2007427} and monitoring\newline - Digital twin for new insights into disease mechanisms. \\ \hline
\textbf{\distributed} & \raisebox{-\totalheight}{\includegraphics[width=7.5cm]{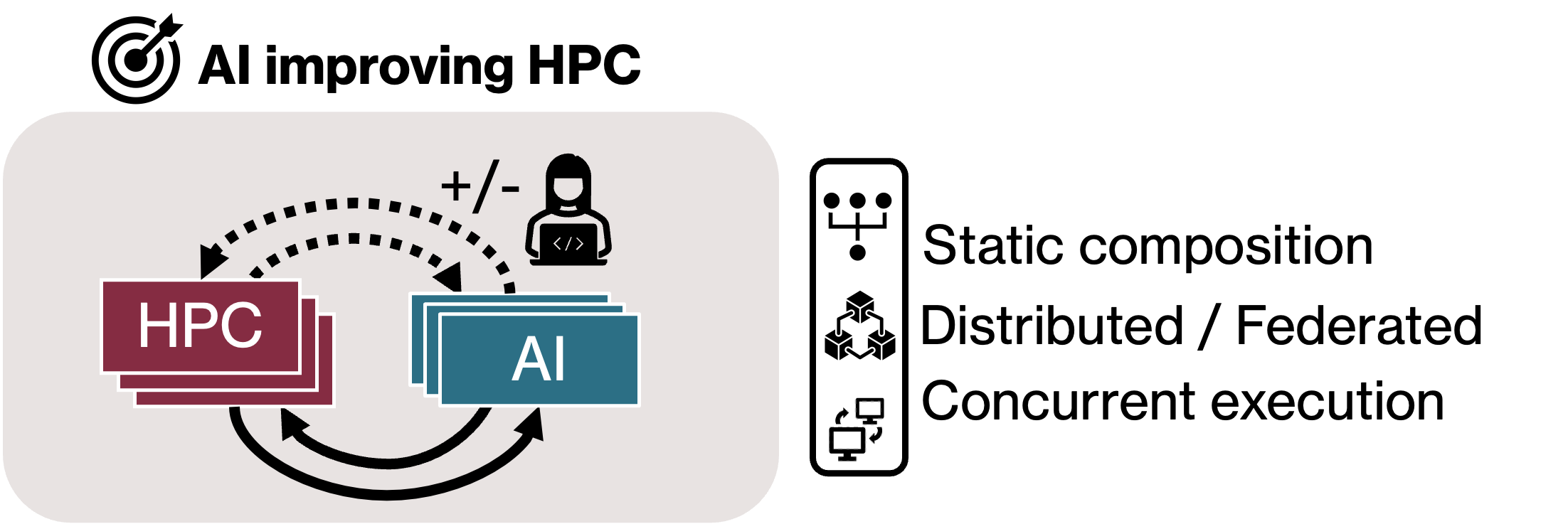}}
                   & 
                   - Edge-to-HPC execution of large-scale simulations\newline
                   - Interactive notebook-based analyses of instrument data and AI output \\ \hline
\textbf{Adaptive\linebreak Training} & \raisebox{-\totalheight}{\includegraphics[width=7.5cm]{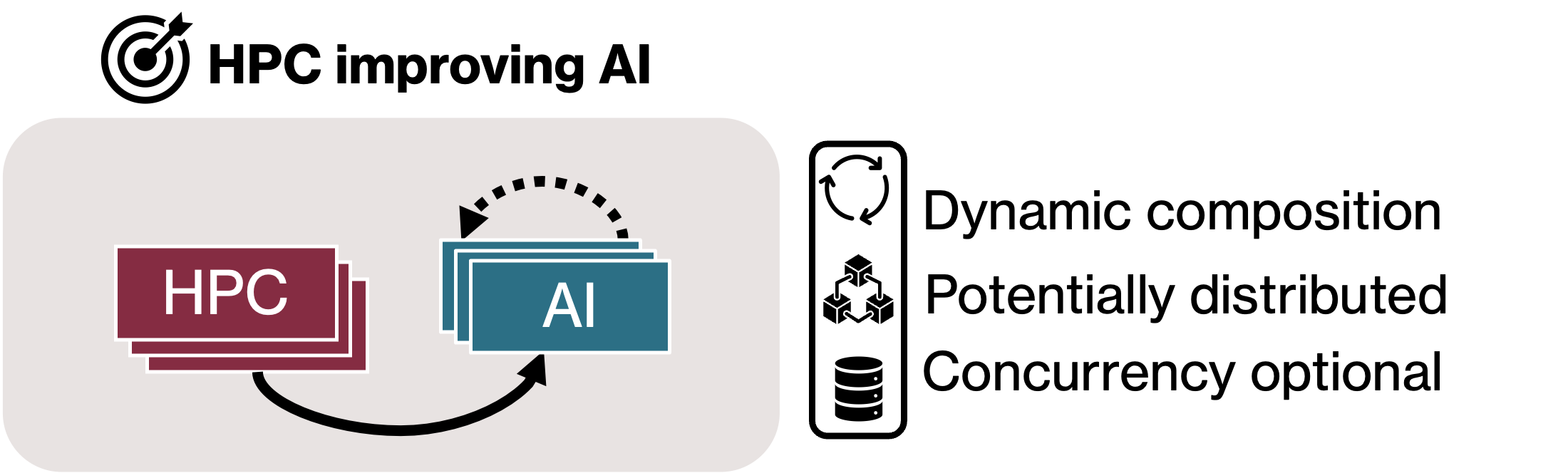}}
                   & 
                   - Hyperparameter optimization workflows\newline - Distributed training of large language models \\ \hline
\end{tabular}
}
\caption{Summary of motif characteristics according to their scope, interaction patterns, and coupling patterns and example use cases. 
}
\label{table:motifs}
\end{table}

\noindent Hereafter, we describe each motif in greater detail and  provide examples of existing and
future implementations of these execution motifs.

\subsection*{\steeringTitle}

\begin{figure}[hbtp]
     \centering
     \begin{subfigure}[b]{0.5\textwidth}
         \centering
         \includegraphics[width=\textwidth]{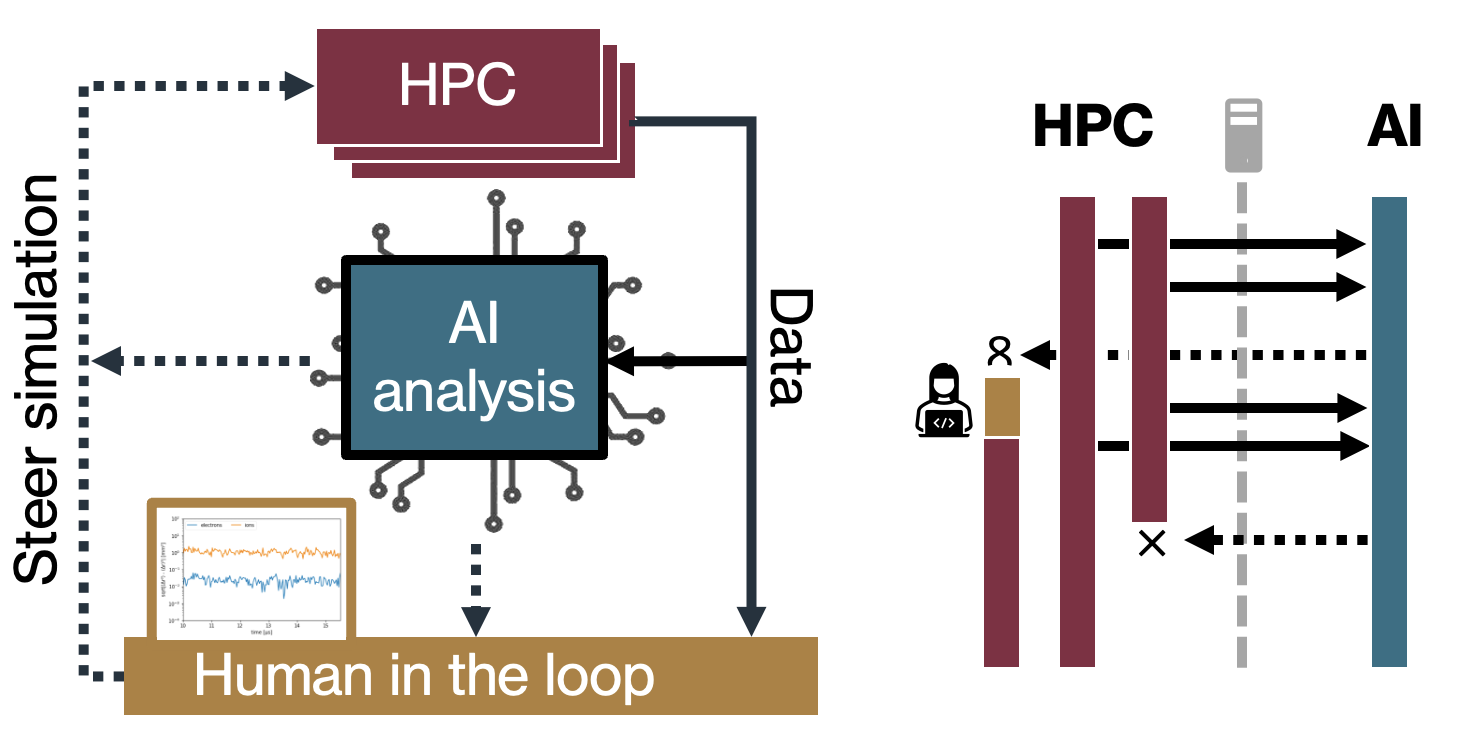}
         \caption{AI system steering (ensembles of) high-fidelity HPC simulations
         to avoid undesired results or guide the workflow towards better solutions.}
         \label{fig:steering}
     \end{subfigure}
     \hspace{0.25in}
     \begin{subfigure}[b]{0.43\textwidth}
         \centering
         \includegraphics[width=\textwidth]{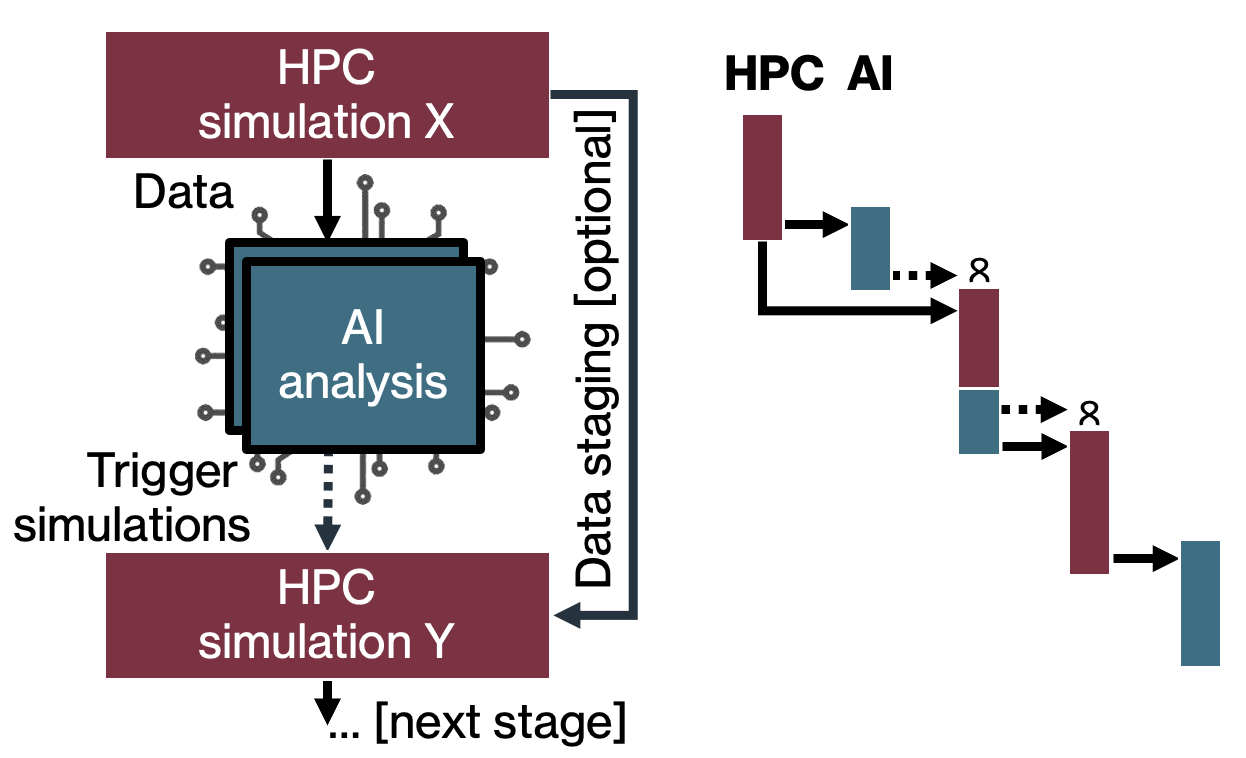}
         \caption{Sequential coupling of data-dependent multistage AI-coupled HPC workflow as a pipeline with filtering/analysis codes between stages.}
         \label{fig:multistage}
     \end{subfigure}
     \hfill
        \caption{Workflow computational patterns for the \textbf{\steering} (a) and \textbf{\multistage} (b) motifs}
        \label{fig:motifs}
\end{figure}

AI-based steering is used in many fields to improve the performance of large multi-physics, multi-resolution simulations, experiments, or an ensemble of simulations. Figure~\ref{fig:steering} presents a typical workflow in this motif where the steering decisions are informed by an AI system, external to the simulations, that analyzes the generated data in near-real time or adaptively selects scientifically meaningful simulations or experiments to run next. 
\begin{compactitem}
   
    \item Interaction patterns: Typically, one or many HPC simulations and/or experiments generate data being analyzed by one AI analysis code. This analysis frequently includes visualization for human-in-the-loop decision-making. Data thus flows from the HPC to the AI, while control goes from AI to HPC.
    \item Coupling patterns: Near real-time analysis is typically required but not always necessary, with some analysis codes steering the configuration of the next ensemble of simulations. Simulations can be terminated, or new instances can be spawned based on the analysis. If experiments are involved, the AI and HPC components might be executed in a distributed fashion. However, such workflows are typically executed on a single system.
    \item Scope: Workflows in this motif usually create models of the simulations offline, based on previous results or synthetic data, 
    and use the AI inference online to improve the HPC components.
\end{compactitem}

The following literature survey focuses on workflows for the {\it \steering} execution motif. These workflows use AI systems to analyze how ``well'' (ensembles of) simulations are progressing toward their initial scientific objective. Based on that analysis, they can change the ``trajectory'' when exploring the parameter space. This allows them to avoid undesired results either or increase performance or accuracy.
The training of the AI system is done offline before the execution of the workflow. Workflows in which the steering process is also used to guide the online training of the AI system will be investigated as part of the {\it \adaptive} motif.

The AI-based steering execution motifs can be found in workflows designed to solve scientific problems in molecular dynamics~\cite{brace2022coupling, 10.1145/3592979.3593420}, fusion~\cite{9297046}, or energy systems simulations~\cite{8339087}. These workflows show orders of magnitude speed-ups in pursuing their scientific objective. For example, this speed-up can be measured as the gain in terms of simulated time performed while covering the same parameter space, quantified by the states sampled by the HPC simulation. The corresponding studies use either custom-designed solutions to couple the codes~\cite{10.1145/3592979.3593420} or workflow management systems, such as DeepDriveMD~\cite{DBLP:journals/corr/abs-1909-07817} for protein folding simulations, or Colmena for electrolyte design~\cite{ward2021colmena}.

Generally, the steering is done in near real-time, with the AI inference running concurrently and interacting with the simulations. A  human-in-the-loop component can be added to the workflow, as in~\cite{DBLP:journals/corr/abs-1807-00567}, to enable the interactive deployment and steering of ensembles of fluid dynamics simulations. This framework provides distributed steering and visualization through a user interface. It allows for the performant execution of interactive simulations for many scenarios while increasing their accuracy via hierarchical refinements thanks to inputs from both the analysis codes and scientists.

DeepDriveMD~\cite{brace2022coupling} is a general-purpose framework to run an analysis code (as a separate application) that pre-processes the simulation data (e.g. selecting only a subset of atoms of interest or calculating physical parameters for a protein's native state), and then uses machine learning (ML) models on the pre-processed data to decide which simulations to run next and terminates the less productive Molecular Dynamics (MD) simulations. By focusing on promising regions of the parameter space, AI-based steering can efficiently navigate complex, high-dimensional systems that were previously intractable. AI systems can also significantly reduce the computational resources required to obtain accurate results by steering simulations toward more relevant configurations. 

The aforementioned example is a typical implementation of the \steering motif, but other examples using different coupling solutions while sharing the characteristics presented in Table~\ref{table:motifs} exist. The AI system analyzes the output of the HPC simulation and either steers the existing simulation or triggers new simulations based on the outcome.

\subsection*{\multistageTitle}

Many HPC applications are executed as a pipeline with distinct stages and data dependencies between these stages. The transition between subsequent stages can be controlled via AI-based logic that decides whether the next stage needs to be triggered based on reaching some set criteria or meeting a specified objective. A transition can also involve an analysis to filter out data between subsequent stages. All these patterns are composed into the \multistage motif illustrated by Figure~\ref{fig:multistage}.

\begin{itemize}
    \item Interaction patterns: Data moves from HPC stages generating data to one or many AI analyses consuming it, but can also be passed between HPC applications executed in sequential stages. AI controls which HPC stage will be executed. Thus, data and control flow in one direction in a many-to-many or many-to-one fashion. Most of the time, the process is automated without requiring humans to analyze the output of the AI codes.
    \item Coupling patterns: While near real-time interaction is a requirement, processes are not typically spawned or terminated as in the Steering motif. However, the workflow flow is dynamic by definition, with branching based on filters. Typically, the entire workflow is executed on a single system.
    \item Scope: AI is primarily used to branch the execution of HPC to guide it toward optimized results. The scope of the \multistage motif is thus to improve HPC. 
\end{itemize}

The following examples are typical implementations of the \multistage motif that we identified in the literature. The AI/HPC stages are either separate applications or integrated into multiple hybrid applications typically deployed on HPC systems sequentially or pipelined.

High-Throughput Virtual Screening Pipelines are emblematic of this motif. For instance, a virtual drug discovery pipeline can use AI to improve the effective sampling of individual stages~\cite{saadi2020impeccable}. 
Integrating multiple \revised{AI-coupled workflows} into a single multistage pipeline is another example of AI-driven discovery~\cite{doi:10.1177/10943420221128233}.
AI can also be used in workflows composing multiple multi-fidelity simulations to dynamically adjust how many and what fidelity levels to run based on the desired outcome~\cite{woo2022optimal}.

Less complex multistage pipelines can be found in neuroscience~\cite{Esteban2019, Huo2016-cd} and cancer research~\cite{Foran2022-ms}. Both fields employ workflows composed of a sequence of medical image processing applications that process the same Magnetic Resonance Image (MRI) or Whole Slide Image (WSI) with the output of one application informing the following decision. Each image-processing application is typically a mix of HPC and AI codes composed of multiple stages, including cleaning, standardizing, analysis, and post-processing within one large code or as separate applications. One such workflow is, for instance, used to identify lesions in the brain~\cite{Huo2016-cd}. It consists of a 3-stage pipeline, starting with a set of MRI images that go through a whole segmentation code, followed by a surface reconstruction application, and finally by an application to create a structure of the brain in MRI space. Based on the output of these three applications, the workflow can further trigger additional codes depending on the research objectives. 

Similar strategies are used in biological and material sciences~\cite{pollice2021data}. For instance, an AI system can parse the output of a first simulation to extract the respective starting configurations of an ensemble of simulations, each performed on patches consisting of multiple lipids and containing one or more proteins~\cite{https://doi.org/10.1002/wcms.1620}.
With further advancements in AI algorithms in these fields, the coupling patterns might also evolve, allowing a closer interaction between the AI and HPC. This can be done either by having the AI produce new intermediate data on demand or by allowing the AI to integrate data from various resources to optimize existing ones in the HPC simulations that are executed concurrently.

\subsection*{\inverseTitle}

Deep learning has been used to solve inverse problems, i.e., determining causal factors from a set of observations, in many scientific fields, including material science~\cite{ma15051811, Jain_2014}, nanophotonics~\cite{SoBadloeNohBravo, Pestourie:18, https://doi.org/10.1002/inf2.12116}, or heat transfer~\cite{PhysRevApplied.16.064006}. However, the lack of sufficiently large training data sets is often a crucial issue for these learning algorithms since their accuracy heavily relies on the amount and the quality of the available data. This is, for instance, the case with material characteristics identification, which requires a huge amount of user time to acquire observational data. HPC simulations have recently started generating sufficient noise-free data to train AI models. The new challenge is getting accurate simulation results, often requiring extensive testing and expertise to fine-tune the simulation input parameters. This can also be time-consuming and resource-intensive, especially for
complex simulations since it requires validation over many tests. AI-coupled HPC workflows following the {\it \inverse} motif address this data sparsity problem and reduce the demand for domain-specific knowledge in determining the input parameters of a simulation.

\begin{figure}[tbhp]
     \centering
     \begin{subfigure}[b]{0.45\textwidth}
         \centering
         \includegraphics[width=\textwidth]{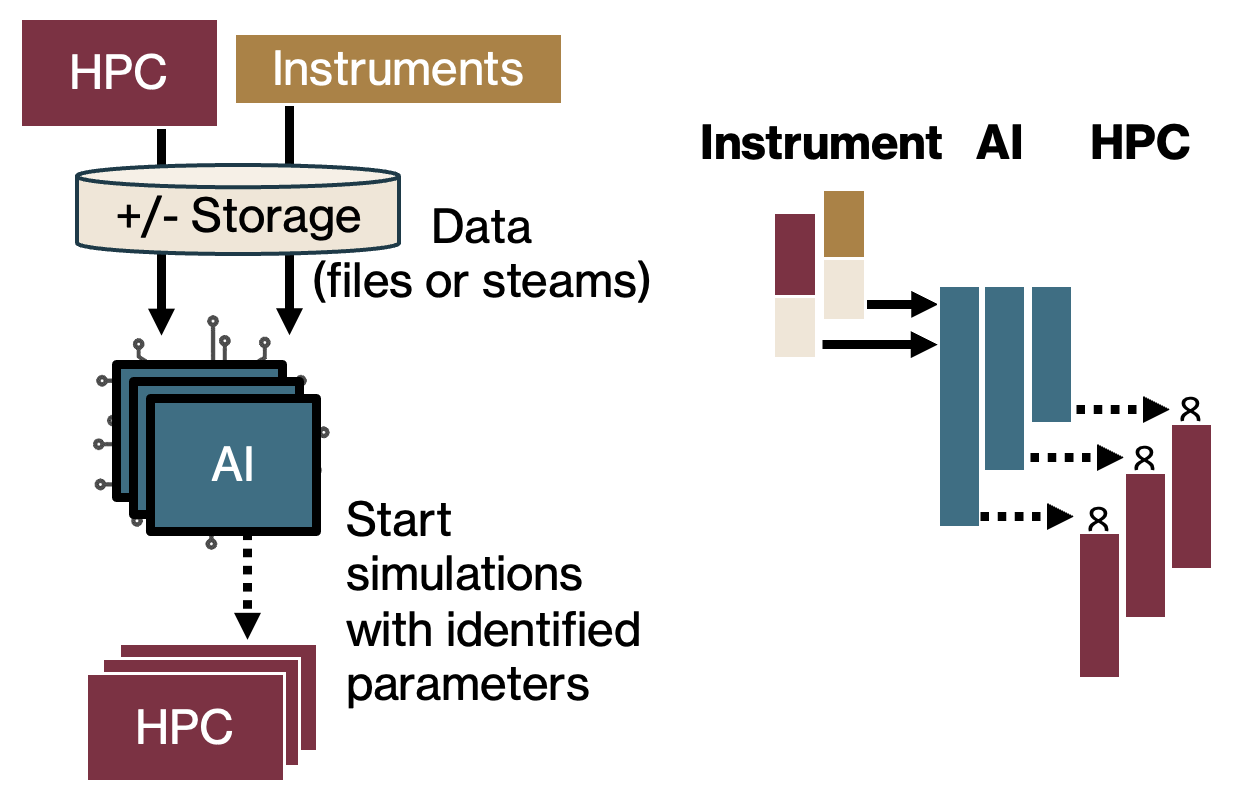}
         \caption{Iterative refinement of design parameters combining experiments with HPC simulations and AI surrogates}
         \label{fig:inverse}
     \end{subfigure}
     \hspace{0.25in}
     \begin{subfigure}[b]{0.45\textwidth}
         \centering
         \includegraphics[width=\textwidth]{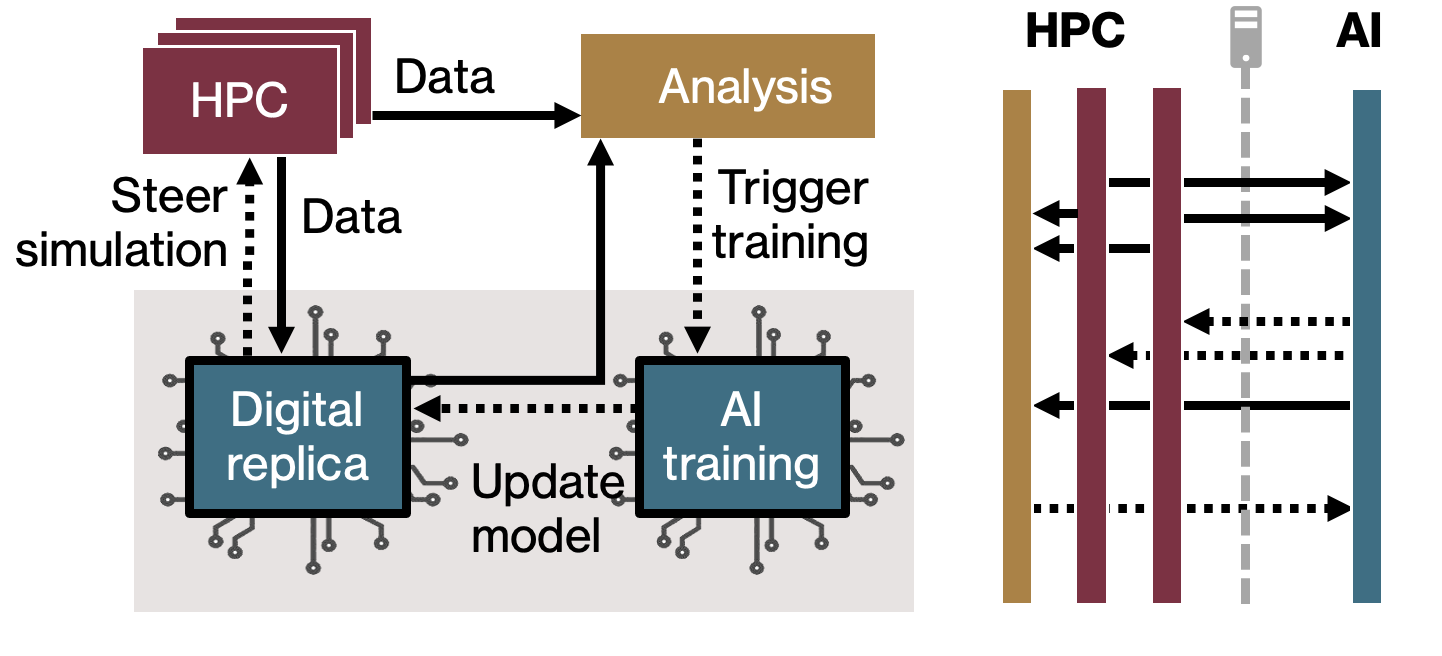}
         \caption{Physics-based models combining HPC high-fidelity simulations and AI reduced order models and/or digital twins}
         \label{fig:digital}
     \end{subfigure}
     \hfill
        \caption{Workflow computational patterns for the \textbf{\inverse} (a) and \textbf{\replica} (b) motifs}
        \label{fig:motifs2}
\end{figure}

Figure~\ref{fig:inverse}  illustrates the typical structure of the \inverse execution motif. \inverse algorithms usually rely on gradient-based AI algorithms searching through a high-dimensional design space's enormous degrees of freedom. They are coupled with simulations or experiments to guide the simulation towards more efficient designs and may re-train or update the models online as the simulation and experiment progress.

\begin{itemize}
    \item Interaction patterns: Workflows in this motif execute multiple HPC simulations (potentially together with instruments), sending data to AI (possibly multiple if online training is used). Control flows from AI to HPC using iterative refinement of design parameters. Typically, there is no human steering in the loop.
    \item Coupling patterns: The AI can use data that has been previously generated (in which case real-time interaction is optional). The execution of the HPC and AI components is static, concurrent, or asynchronous and usually occurs in the same system.
    \item Scope: AI models are improved to reflect physical phenomena better, while the model can improve HPC simulations. Thus, the \inverse motif aims at improving both AI and HPC components. 
\end{itemize}

A powerful illustration of this motif is its application to material design~\cite{9556083}. The method determines material behaviors and consists of
three phases: data generation, training, and the inverse phase. The data generation phase uses a set of HPC simulations, each testing different material properties using various material parameters as inputs. In the training phase, a Deep Neural Network (DNN) model is trained on the simulation outputs and the material parameters. The inverse phase compares the prediction and real input values to improve the DNN model and simulation performance. Based on the model's output, new simulations are triggered, improving both the HPC simulation's performance and the AI system's accuracy.

Examples of inverse design workflows can be found in several other fields: bioinformatics, biochemistry, material science, and nanophotonics.
Recent breakthroughs in AI-based protein structure prediction (i.e., sequence from the structure) have set the stage for the use of AI models for solving the inverse problem in molecular sciences (e.g., protein design)~\cite{stach2021autonomous,nigam2022tartarus,wang2023IPDPS}. Similarly, in materials science~\cite{ZHANG2021109213, doi:10.1021/acsphotonics.2c00968} workflows for the inverse design problem have been implemented where the physics simulation is coupled with different optimization methods (e.g., directional Gaussian smoothing for nanophotonics~\cite{ZHANG2021109213}) to improve the model used. 

\subsection*{\replicaTitle}

The \replica motif incorporates workflows that couple the training or usage of AI systems that simulate the behavior of physical phenomena (e.g., for digital twins~\cite{national2023foundational}) with the HPC simulation and/or a scientific instrument for a
variety of scenarios and situations in a range of disciplines and fields. These workflows typically utilize low-fidelity models or advanced
data-driven AI models to generate insights not possible using traditional
high-fidelity simulations or observational models (given time and computing resource restrictions).
Figure~\ref{fig:digital} presents an overview of the computational
pattern for this motif. The digital replica receives data from the HPC
simulation/instrument. It triggers analyses to steer the
simulation (in near real-time automatically or post-mortem using a
human-in-the-loop intervention to tune the parameters of the next simulation
run). Initial training is typically done offline before the execution of the workflow
and re-triggered to update the model based on discrepancies
between what is predicted by the simulation and feedback from the
analysis. 

\begin{itemize}
    \item Interaction patterns: Data and control flow in both directions, with the HPC simulation/instrument data being used to steer the training of AI models and the data generated by the digital replica being used to predict and control the HPC components. Humans may be involved in monitoring and adjusting current and future runs. 
    \item Coupling patterns: Real-time interaction is required. Typically, the composition of the workflow is static, and the flow patterns change only when visualization is turned on. The AI and HPC components can run in the same facility or are geographically distributed (e.g. when instruments are involved).
    \item Scope: Both AI and HPC components are improved, the AI predictions improve the run of the HPC and the data generated by HPC improve the digital replica.
\end{itemize}

Relevant studies that fit this motif come from diverse domains (e.g., bioinformatics / medical domain~\cite{alber2019integrating, GILLETTE2021102080}) where machine learning and multiscale modeling are combined to create a digital twin, steer expensive analysis codes towards new insights into disease mechanisms, and help to identify new targets and treatment strategies. 

The coupling of HPC simulations with surrogate AI models can be implemented and deployed in HPC systems differently with different performance implications~\cite{9835386}. This study explores three coupling schemes: tightly coupled, which forms a single executable with one-to-one node mapping; semi-tightly coupled, with separate executables mapped in various ways; and loosely coupled, where separate executables run on different machines using many-to-many mapping. It also recommends running digital twins on different architectures and with different models. It shows the potential benefits of these workflows with a use case in materials science.

Fusion energy researchers are turning to digital twins as a promising avenue
for efficient whole-device optimization and virtual component qualification,
crucial for successfully designing fusion reactor components under
conditions that defy real-world testing. Digital twin
workflows are primed to play a crucial role in ensuring the success of \revised{the international thermonuclear experimental reactor}~(ITER)~\cite{HOLTKAMP2007427}, and pave the way for the future of
commercial fusion energy. Several studies~\cite{en16135107, Kropaczek2023,
TINDALL2023113773} presents the vision of the fusion science community to develop future workflows. They all discuss the design of a workflow combining digital twins of different parts of the fusion reactor (e.g., the Central Solenoid Converter Power Supply grid~\cite{en16135107}) coupled with real-time simulation strategies to monitor and study the performance of the ITER grid while the reactor is functional.

\subsection*{\distributedTitle}

\begin{figure}[tbhp]
     \centering
     \begin{subfigure}[b]{0.45\textwidth}
         \centering
         \includegraphics[width=\textwidth]{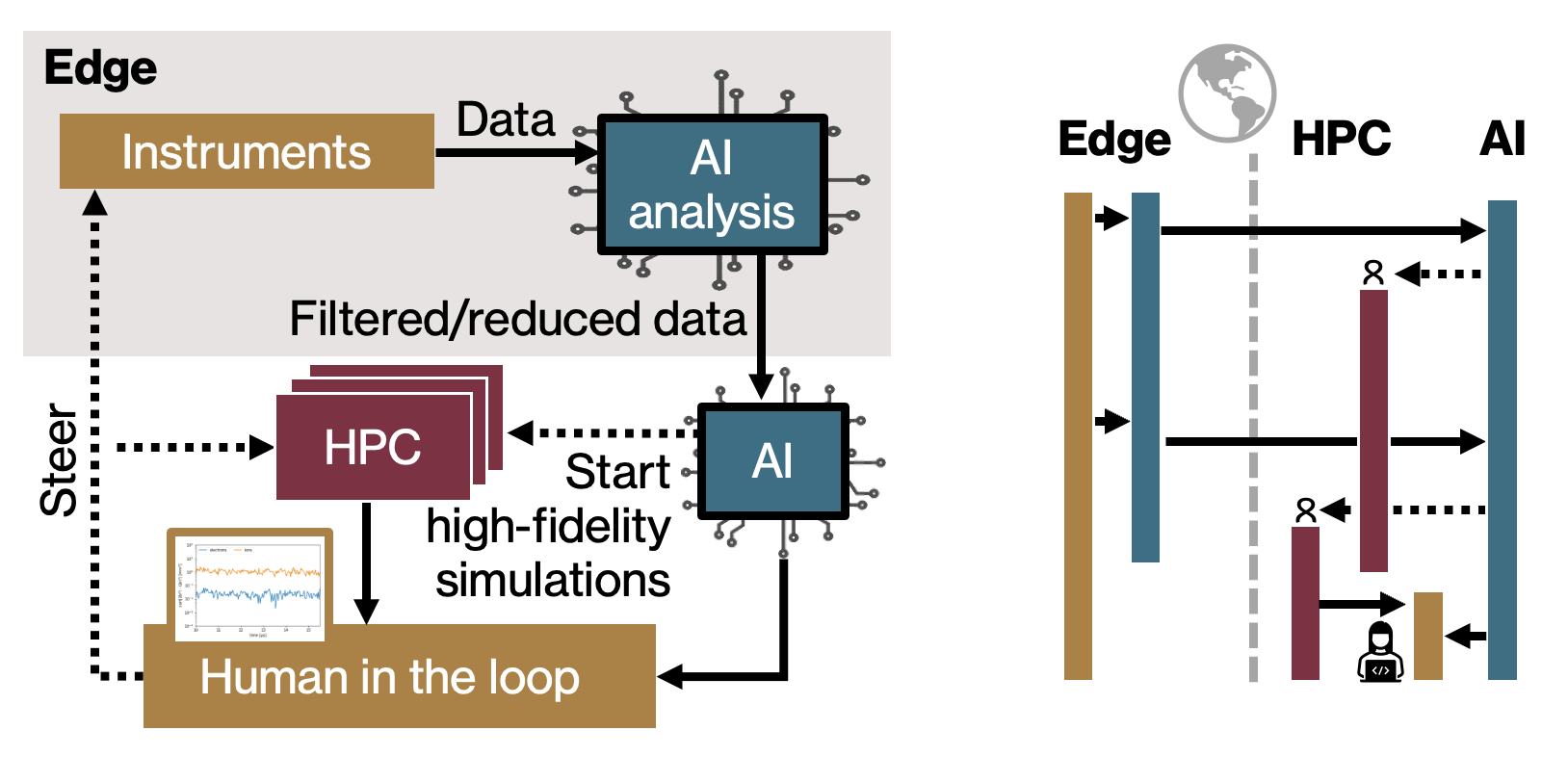}
         \caption{HPC simulations coupled with experiments and AI surrogate models and AI analytics geographically distributed using dynamic interaction and resource management}
         \label{fig:distributed}
     \end{subfigure}
     \hspace{0.25in}
     \begin{subfigure}[b]{0.45\textwidth}
         \centering
         \includegraphics[width=\textwidth]{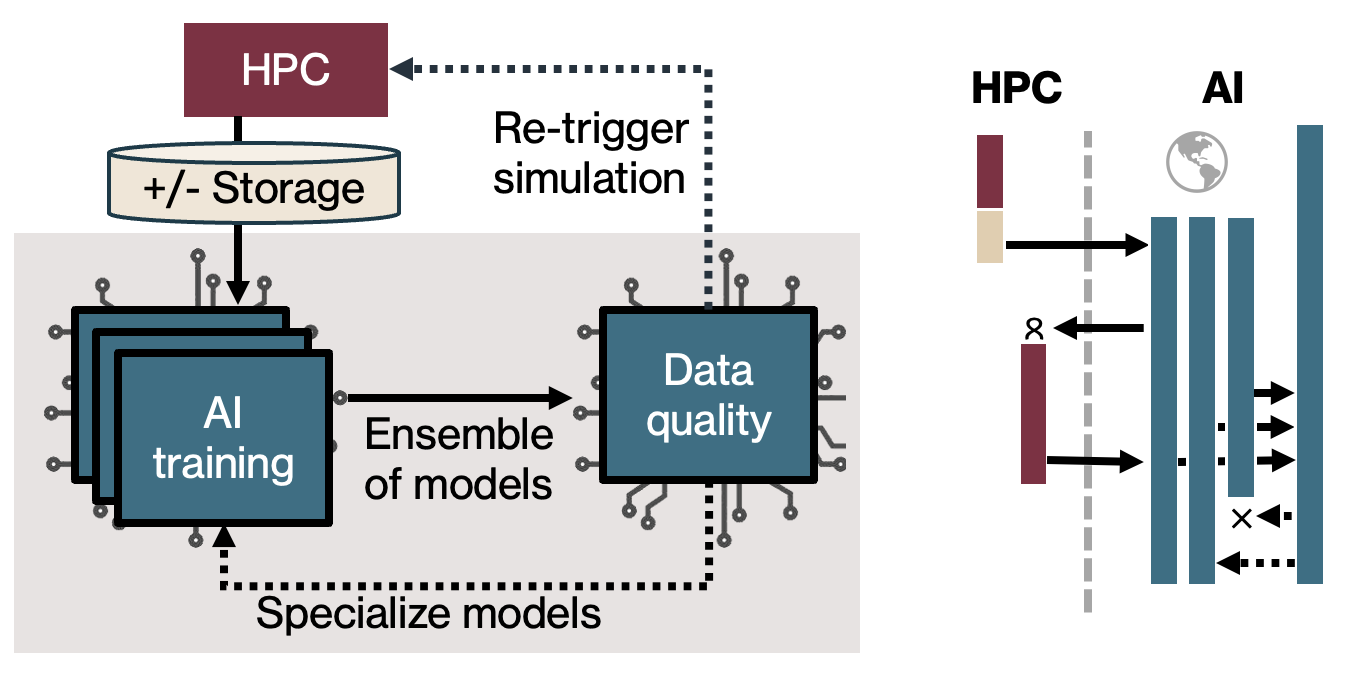}
         \caption{Adaptive methods for training ensemble of models with large datasets on HPC potentially coupled with simulations that
         generate the data needed for training}
         \label{fig:training}
     \end{subfigure}
     \hfill
        \caption{Workflow computational patterns for the \textbf{\distributed} (a) and \textbf{\adaptive} (b) motifs}
        \label{fig:motifs3}
\end{figure}

The data analytics modules (e.g., the AI system) in previous motifs are typically executed using the same computing resources as the simulation. In some cases (e.g., when visualization is used or instrument data needs to be managed), parts of the workflow can be executed on a remote site, including the scientist's laptop. However, resource management strategies are usually static, and workflows do not adapt their structure to performance changes. The rapid development of edge infrastructures and localizing large amounts of potentially unnecessary data have motivated the development of edge analytics. The corresponding execution motif is illustrated by Figure~\ref{fig:distributed}. 
Workflows in this motif couple HPC simulations and instruments with geographically distributed AI digital twins, AI surrogate models, and  AI analytics (including multiple clusters, compute resources on the edge, user laptops, etc.). The workflows dynamically adapt to the network and compute capabilities from the point of view of task placement and resource management (e.g., triggering new analysis or compression, adapting the accuracy of the data transfer, allocating new resources, or spawning redundant computation for better resilience).

\begin{itemize}
\item Interaction patterns: Data and control flow in both directions from multiple AI to multiple HPC (e.g., edge AI monitoring or AI filtering steering, AI digital replicas generating data, HPC simulations generating data, and updating the AI edge analysis components). For near real-time experiment steering, no human is involved in the process; however, configuring the execution of future experiments or large-scale simulations typically requires humans in the loop to monitor the output of the previous analysis.
\item Coupling patterns: Near real-time is required to analyze the state of the simulations/digital replicas and the state of the networks for the workflow to adapt. The workflows typically use static composition in the executed components but employ dynamic flow paths and configurations to provide good performance regardless of the state of the systems and applications. The execution is always performed in a federated manner.
\item Scope: The main scope of workflows in the \distributed motif is to improve the execution and/or productivity of the HPC simulation or instruments.
\end{itemize}

Typical examples of this motif include diverse models on the edge-to-exascale continuum~\cite{rosendo2022distributed} and coupled simulations across multiple HPC centers~\cite{KHAN2019398, https://doi.org/10.1002/pro.3721}. The workflows presented in~\cite{rosendo2022distributed} combine resources and services at the center of the network (in Cloud and HPC data centers) with computing and storage capabilities at the edge and in transit along the data path. 

For coupling simulations across multiple HPC centers, it is possible to rely on a single workspace that provides a global view of information shared from multiple geo-distributed HPC data centers~\cite{KHAN2019398}. The feasibility of distributed workflows has been shown by coupling the \revised{scientific tools for macromolecular modeling and protein design} with interactive notebook-based analyses executed outside the HPC center~\cite{https://doi.org/10.1002/pro.3721}. Another application of this approach is for neutron scattering science~\cite{yin2023toward}, for which the performance challenges related to the transfer of large files between the HPC center and the experimental facility have been identified as one of the main performance bottlenecks for the deployment of workflow motif. A complementary work to these studies investigates the overhead and performance of coupling non-HPC systems (equivalent to edge devices) to handle HPC workflows~\cite{doi:10.1177/00375497211064198}. 

Finally, a survey of the field of federated learning~\cite{prigent2022supporting} highlights potential performance bottlenecks. The study of challenges associated with orchestrating complex workflows in geographically dispersed environments~\cite{doi:10.1177/1094342018778123} also highlights the lack of efficient mechanisms for managing multistage workflows that dynamically traverse distributed computing resources, edge devices, and instrument networks. Addressing this gap requires novel paradigms for orchestrating workflow execution across diverse platforms and geographical boundaries.

\subsection*{\adaptiveTitle}\label{sec:training}

The previous motifs include training as part of their execution workflow either offline before running the workflow or online by updating the models based on the discrepancies between the simulation output and the predictions. This last motif (\adaptive) focuses on the training phase as a standalone workflow. It encapsulates all the adaptive methods to train one or an ensemble of models using an existing dataset or running coupled with the HPC simulation that generates the required training data (Figure~\ref{fig:training}). The training algorithm can interact with the simulation and guide its execution to guarantee better data coverage for building robust models. This guidance can be done within the training code or through a separate analysis code (e.g., the figure's ``Data quality'' box). The analysis code that verifies the accuracy of the created models may trigger a new training event (e.g., using different model parameters) or spawn new simulation instances. The workflow can have a human-in-the-loop component to guide the process towards specialized models.

\begin{itemize}
\item Interaction patterns: Data flows from HPC to multiple AI; if surrogates are used to generate data, the data flow is from AI to AI training. Control flows from the analysis code to the HPC and AI training. Typically, the process is automatic, and the human controls only the initial configuration of the workflow.
\item Coupling patterns: Real-time coupling between AI and HPC is optional since previously generated datasets can be used. The AI and analysis components require real-time coupling. Dynamic composition is required, with the analysis code optimizing the training process by terminating low-accuracy models and spawning processes with configurations likely to increase the accuracy. Typically, the training is done on the same system. However, federating learning might start using similar workflows in the future.
\item Scope: The scope of the \adaptive workflows is to train accurate models and to improve AI components.
\end{itemize}

\revised{
There have been numerous studies applying Hyperparameter Optimization (HPO) techniques--including Bayesian optimization~\cite{balaprakash2018deephyper}, evolutionary strategies~\cite{DBLP:journals/corr/abs-1909-12291}, and genetic algorithms~\cite{8638645}--to improve surrogate models and reduced-order representations for HPC applications. These approaches have been used across domains such as cancer pathology~\cite{DBLP:journals/corr/abs-1909-12291}, spacecraft landing control~\cite{9762951}, and molecular simulations. Similarly, Neural Architecture Search (NAS)~\cite{geifman2019deep, marchisio2022rohnas} and Large Language Model (LLM) training~\cite{barham2022pathways, chowdhery2022palm, blanchard2022adaptive} increasingly leverage simulation-aware feedback loops to explore model configurations more efficiently.
Dynamic coupling of these learning agents with HPC simulations--whether automated or human-guided--is gaining traction in fields such as fusion~\cite{9297046}, fluid dynamics~\cite{MAO2023102171}, energy systems~\cite{8339087}, and climate science~\cite{10.1145/3624062.3624283}. The FASTRAN-IPS workflow~\cite{9297046} exemplifies this approach by training a surrogate model iteratively while steering simulations to explore undersampled regions of the parameter space. AI is also paired with visualization systems to allow users to interactively explore results and launch new simulations as needed~\cite{8339087}.
}

\subsection*{Summary: Scientific Workflow Applications Using Multiple Motifs}

The motifs presented in this section use different coupling and interaction patterns when executing HPC and AI for different scopes. These motifs are not mutually exclusive; we expect future large-scale workflows to combine several motifs. One such example deploys an AI-\revised{coupled HPC} workflow, combining three motifs~\cite{casalino2020aidriven}: (1) it adaptively \textit{steer} MD simulation ensembles; (2) it further couples this workflow with AI-based methods in a \textit{multistage pipeline} fashion; (3) it uses \textit{adaptive execution} to learn which parts of the parameter space have been sampled sufficiently and to trigger simulations from under-sampled regions. Each component depends on a distinct solution: the steering uses DeepDriveMD, the learning employs a 3D PointNet-based adversarial autoencoder~\cite{qi2017pointnet}, and the overall integration relies on a custom solution. 

\revised{Scientific computing is evolving towards integrated research infrastructures (like IRI~\cite{brown2023integrated})} 
that will unite data resources, experimental user facilities, and advanced computing resources to accelerate the deployment of large-scale simulations and workflows. Such integrated infrastructures will accelerate the trend of 
coupling multiple HPC simulations, surrogate models, and digital twins with AI analysis 
at multiple levels. 

However, while initiatives like the IRI will drastically simplify workflow building and deployment, enabling the efficient execution of complex workflows that span diverse scientific landscapes and combine multiple AI-HPC coupling motifs is non trivial. There is currently no solution that fits all requirements across multiple applications and fields and often specialized solutions are preferred at all levels of the software stack.  
In the next sections, we present solutions for workflow management systems targeted at individual motifs and/or specific to certain domains, discuss the performance of current workflows, and highlight potential directions for this fast-paced changing field.

\section{AI-coupled HPC Workflow Management Frameworks}\label{sec:frameworks}

For two decades, automation, simplification, and workflow management optimization have been active research topics in the HPC and distributed systems communities. This has led to developing many libraries, frameworks, and tools for domain scientists to advance their scientific agenda. More recently, diverse research communities in both industry and academia have been working on allowing AI frameworks to go beyond being a standalone black-box operator to facilitate their inclusion into more complex workflows. While the previous section highlighted that several studies involving AI-coupled HPC workflows relied on custom solutions with low portability and high maintenance cost, we observe a convergence of the efforts related to workflow management in both the HPC and AI communities with the recent development of workflow frameworks specifically targeting \revised{such} workflows. We review some of these frameworks and identify their connections with the execution motifs identified in the previous section. Then, we identify several challenges related to the development of those frameworks.

\subsection{Overview of Workflow Management Frameworks}
\label{sec:framework_overview}

\revised{To complement the execution motifs introduced earlier, Table~\ref{tab:aihpc_tools} summarizes key frameworks developed for AI-coupled HPC workflows. Some are domain-specific, while others offer general-purpose capabilities. In the discussion below, we cluster frameworks with similar architectural patterns and highlight representative use cases from each category.}

\begin{table}[hbtp]
\centering
\begin{tabular}{|p{3cm}|p{4.2cm}|p{4.4cm}|p{2.7cm}|}
\hline
\textbf{Framework} & \textbf{Application Domain} & \textbf{Execution Motif(s)} & \textbf{Underlying System} \\ \hline
CANDLE~\cite{10.1145/3337821.3337905, candle} & Cancer research & \steering | \multistage | \adaptive & \centerline{SWIFT/T} \\ \hline
DeepDriveMD~\cite{deepdrivemd} & Biophysics/Molecular \linebreak Simulations & \steering | \multistage | \adaptive & \centerline{Radical} \\ \hline
MuMMI~\cite{mummi1} & Biophysics/Molecular \linebreak Simulations & \steering | \multistage | \adaptive & \centerline{Maestro~\cite{maestrowf}}\\ \hline
IMPECCABLE~\cite{saadi2021impeccable} & Virtual drug discovery pipeline & \multistage |  \inverse | \adaptive & \centerline{Radical} \\ \hline \hline
Colmena~\cite{ward2021colmena} & General-purpose steering of \hfill \hfill \linebreak ensembles & \steering | \adaptive & \centerline{Parsl} \\ \hline
\revised{EXARL\cite{exarl-github, EXARL-PARS}} & \revised{General-purpose\linebreak Reinforcement Learning control} & \revised{\steering | \adaptive} & \revised{\centerline{MPI}} \\ \hline
\revised{Globus\linebreak Compute~\cite{chard2020funcx, li2022funcx}} & \revised{Function-as-a-Service on federated resources framework} & \revised{\multistage | \distributed} & \revised{\centerline{Globus, Parsl}} \\ \hline
\revised{INTERSECT~\cite{engelmann2022intersect}} & \revised{Interconnected Science\hfill\hfill \linebreak Ecosystem} & \revised{\distributed | \steering | \multistage | \replica} & \revised{\centerline{-}} \\ \hline
SmartSim~\cite{balin2023situ, partee2022using}  & A scalable, open-source, multi-physics simulation API. & \steering | \multistage | \replica | \adaptive & \centerline{Redis} \\ \hline
Stimulus~\cite{stimulus} & General-purpose data\hfill\hfill\linebreak management library & \inverse | \replica | \adaptive & \centerline{-}\\ \hline
NoPFS~\cite{10.1145/3458817.3476181} & General-purpose I/O \hfill\hfill\linebreak middleware for model training& \inverse | \replica | \hfill\hfill\linebreak \adaptive& \centerline{-}\\ \hline
\end{tabular}
\caption{Characteristics of domain-specific and general-purpose AI-coupled HPC Workflow frameworks.}
\label{tab:aihpc_tools}
\end{table}

The \textbf{Cancer-Distributed Learning Environment~(CANDLE)}~\cite{10.1145/3337821.3337905, candle}
application workflow includes data acquisition, 
analysis, model formulation, and molecular dynamics simulations. AI is used to
develop and test protein binding hypotheses and better understand proteins' behavior on cell membranes. This requires performing
Hyperparameter optimization to identify the most effective model
implementations efficiently. CANDLE directly incorporates parameter exploration methods
for the efficient exploration of parameter spaces of orders greater than
$10^9$. These methods manage large amounts of data and rely on scalable
data parallelism to speed up the AI model.

\revised{
Several domain-specific frameworks--DeepDriveMD, MuMMI, and IMPECCABLE--share a common loop of simulation, AI-based analysis, and adaptive steering. These frameworks often operate on ensembles of simulations, use AI to select promising subregions or filter states, and incorporate feedback loops to optimize sampling or model fidelity.
\textbf{DeepDriveMD}~\cite{deepdrivemd} focuses on protein folding. It uses deep learning to construct latent representations of molecular conformations and steers ensembles of molecular dynamics (MD) simulations accordingly. RADICAL-Cybertools orchestrate simulation workflows and model-driven steering.
\textbf{MuMMI}~\cite{mummi1} couples macro-scale density functional theory simulations with micro-scale MD simulations. An AI-driven importance-sampling engine bridges the two scales, dynamically identifying sub-regions for follow-up simulation. MuMMI employs the Maestro workflow system and in situ data aggregation to manage the high data volume.
\textbf{IMPECCABLE}~\cite{saadi2021impeccable} integrates surrogate modeling and AI-guided sampling into a three-stage pipeline for drug discovery. It uses Radical-Cybertools to manage the workflow across leadership-class systems.
}

\revised{Beyond domain-specific frameworks, a variety of general-purpose workflow systems have been developed to support AI-HPC integration across scientific fields. These frameworks differ in orchestration models, data handling strategies, and support for federated or in situ execution.}

\revised{
\textbf{Colmena}~\cite{ward2021colmena} supports AI-guided steering of simulation ensembles. Built on the Parsl workflow system, Colmena enables concurrent execution of simulations, AI retraining, and decision-making agents. It introduces the ``Thinker'' interface, which coordinates task scheduling using Redis-backed queues and allows for flexible, user-defined control strategies.

\textbf{EXARL}~\cite{exarl-github, EXARL-PARS} enables reinforcement learning at HPC scale by implementing a learner/actor architecture within a parallel MPI-based training loop. It supports concurrent learning and simulation with OpenAI Gym environments and has been integrated with domain-specific workflows such as CANDLE for cancer research.

\textbf{Globus Compute}~\cite{chard2020funcx, li2022funcx} (formerly FuncX) is a federated Function-as-a-Service (FaaS) platform designed for cross-site workflows. It uses a REST-based model to invoke user-defined functions across distributed computing endpoints, including edge devices, clusters, and HPC systems. It has been demonstrated in real-time AI analysis pipelines for experimental data, such as x-ray ptychography.

\textbf{INTERSECT}~\cite{engelmann2022intersect} is a distributed, microservices-based framework developed at Oak Ridge National Laboratory to connect robotic labs, instruments, and compute facilities. It enables self-driving labs through AI-guided experiments and real-time data streaming. The system includes a control message queue and scientific data layer, and has been applied in electrochemistry experiments.

\textbf{SmartSim}~\cite{balin2023situ, partee2022using} is an in situ execution framework that couples simulations with AI models using an in-memory Redis database. It supports data exchange between Fortran/C++ simulation codes and Python-based ML tools via a publish/subscribe model. This design allows low-latency communication without reliance on file-based I/O.

\textbf{Stimulus}~\cite{stimulus} addresses data ingestion challenges by providing a unified interface across scientific data formats (e.g., HDF5, NetCDF, ADIOS) for use in AI frameworks such as PyTorch and TensorFlow. It simplifies access pattern mismatches and decouples HPC simulation output from AI pre-processing code.

\textbf{NoPFS}~\cite{10.1145/3458817.3476181} is a Machine Learning I/O middleware that optimizes data prefetching during model training. By leveraging knowledge of randomized data access patterns, it reduces contention on shared filesystems. Though not a full workflow system, it improves training throughput and indirectly benefits simulation workloads running on the same infrastructure.
}

\revised{AI integration into real-time and streaming workflows extends beyond HPC contexts. For instance, Kafka-ML~\cite{Kafka-ML} connects Apache Kafka with AI models to process data from IoT sensors. Similarly, RedisAI enables inference directly within Redis streams~\cite{partee2022using}. As efforts like the DOE Integrated Research Infrastructure expand, we expect stream processing frameworks to play a central role in AI-coupled HPC workflows.}

\subsection {Challenges}
Studying these different frameworks highlights two main challenges to
developing AI-coupled HPC workflow management frameworks. The first challenge is the coexistence and communication between 
HPC and AI tasks in the same workflow. This communication is mainly impaired by the difference in
programming languages used in HPC (i.e., C, C++, and Fortran) and AI (i.e.,
Python). One solution to circumvent this language barrier would be to consider
that all communications between HPC simulations and AI models are done through
data. However, this raises other issues related to differences in data format
and access patterns~\cite{suter2023escience}. The successful and efficient management of AI-coupled HPC
workflows thus call for a more unified data plane management in which
high-level data abstractions could be exposed, and the complexities of the format
conversion and data storage and transport could be hidden from both the HPC
simulations and AI models.

The second challenge concerns using the insight provided by AI
models in the HPC simulations. For the motifs outlined, this insight is used
to guide, steer, or modify the shape of the workflow by triggering or stopping
new simulations. This implies that the workflow management system must be able
to ingest and react dynamically to inputs coming from the AI models. However,
most existing workflow management systems enforce a static description
and execution of the workflows. New methods must be designed to enable more flexible and adaptive workflow executions.
Moreover, the more complex and evolving data transfer patterns associated with
such more dynamic workflows may directly impact the performance of AI-coupled
HPC workflows if not managed properly. As feedback loops have additional
dependencies between workflow components, any delay in data production and
storage directly impacts the components that consume data, which have to wait
to start their execution and remain idle.

In this section, we considered frameworks developed to automate the execution of AI-\revised{integrated} workflows in different scientific domains or designed to address some common challenges from the perspective of their matching with the proposed execution motifs. However, the sheer amount of data handled by these frameworks (e.g., ''scaling to full memory ($>$ PB) on the largest supercomputers''~\cite{10.1145/3337821.3337905}), the high level of task concurrency they manage (e.g., 16k concurrent tasks for DeepDriveMD~\cite{deepdrivemd}) or both (e.g., 36k concurrent tasks and several PB of raw data for MuMMI~\cite{mummi1}) can have some important performance implications. In the next section, we focus on these performance considerations of \revised{such} workflows. 

\section{Performance}\label{sec:performance}

To analyze the performance of AI-coupled HPC workflow executions, we consider its fundamental components and various strategies. Furthermore, we discuss each workflow component's challenges and performance implications and explore how different
coupling strategies affect performance. Then, we review the six
execution motifs in the context of these components and their coupling, addressing the
various challenges and bottlenecks. Summarizing the section in Table~\ref{table:performance}, we provide a comprehensive overview of the main characteristics of each
motif, the associated typical performance bottlenecks, and coupling
strategies. Lastly, we investigate the implications of software implementation
and the significance of benchmarking such workflows.

As we delve deeper into the intricacies of \revised{AI-integrated} workflows, it becomes evident that several open issues
emerge, particularly regarding interoperability, data handling, and resource
allocation. These challenges underscore the importance of
understanding the performance and highlight broader opportunities to enhance
the efficiency and effectiveness of such workflows.

\subsection{Workflow Components}

Each workflow component has performance implications individually and in their coupling. In general, for AI, especially for AI training, there is a trade-off between
{\em computation} vs. {\em communication} \cite{pati2023computation}.
Offloading to AI accelerators requires data movement and may be largely
affected by how the accelerators are integrated into the main supercomputer. 
GPUs typically achieve lower latencies for large batch queries, although they incur significant loading overhead; conversely, CPUs usually exhibit lower loading latencies and excel with smaller batch sizes \cite{yadwadkar2019case}.  
This topic has been studied comprehensively in Ref.~\cite{IntelligentResolution} for an AI-driven simulation across
cross-facility infrastructure performing simulations on a traditional HPC
supercomputer while offloading the AI model to a Cerebras accelerator, a wafer-scale application-specific integrated circuit (ASIC) device.
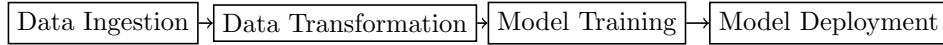
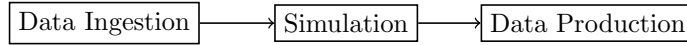
\begin{figure}[hbtp]
     \centering
     \begin{subfigure}[b]{0.9\textwidth}
         \centering
         \begin{tikzpicture}[->, auto, node distance=3.2cm, semithick]

\node [draw] (A) {Data Ingestion};
\node [draw, right of= A] (B) {Data Transformation};
\node [draw, right of= B] (C) {Model Training};
\node [draw, right of= C] (D) {Model Deployment};

\draw [->] (A) -- (B);
\draw [->] (B) -- (C);
\draw [->] (C) -- (D);
  
\end{tikzpicture}
         \caption{Machine learning training and deployment pipeline}
         \label{fig:ai-pipeline}
     \end{subfigure}
     \vspace{0.2in} \\
     \begin{subfigure}[b]{0.9\textwidth}
         \centering
         \begin{tikzpicture}[->, auto, node distance=3.2cm, semithick]

\node [draw] (A) {Data Ingestion};
\node [draw, right of= A] (B) {Simulation};
\node [draw, right of= B] (C) {Data Production};

\draw [->] (A) -- (B);
\draw [->] (B) -- (C);
  
\end{tikzpicture}
         \caption{Pipeline for typical HPC applications}
         \label{fig:hpc-pipeline}
     \end{subfigure}
     \hfill
        \caption{The HPC and AI pipelines concurrently running within a workflow typically overlap different stages depending on the type of workflow they are deployed in}
        \label{fig:workflow-pipeline}
\end{figure}

Understanding these performance implications is crucial when considering the entire machine learning training and deployment pipeline, as shown in Figure~\ref{fig:ai-pipeline} \cite{baylor2017tfx}.
The computation and communication trade-offs inherent in using accelerators like GPUs or specialized hardware such as Cerebras influence this pipeline at each step. For example, while GPUs might accelerate the Model Training stage for large batch sizes, the initial Data Ingestion and Data Transformation stages might still depend heavily on traditional CPU processing. This interplay between different hardware capabilities emphasizes the need for a nuanced approach in optimizing each pipeline stage for efficient AI model development and deployment in HPC environments. Note that we do not consider the evaluation of the model in
performance analysis, as it is primarily concerned with the accuracy of the model, while its computational cost will be discussed in the section on the Model Deployment.

The HPC pipeline illustrated in Figure~\ref{fig:hpc-pipeline} has typical I/O and communication network performance considerations. As HPC simulations evolved to include GPU kernels and/or AI in their codes, the trade-offs described in the previous paragraph are also becoming relevant for the HPC components. When deployed together, the AI and HPC components will overlap different stages of their respective pipelines and face new performance bottlenecks that are unnecessary during individual runs. For example, the Data Production phase of the HPC application will typically overlap with Data Ingestion for the \steering motif, which might cause concurrent read/write access to the same datasets, resulting in performance penalties due to I/O contention. 

\subsubsection{Data Ingestion} 
Traditional HPC simulations rely on I/O libraries highly tuned towards large sequential file access (like ADIOS2 or HDF5), Deep learning frameworks typically train
on mini-batches of data, which tends to perform poorly on parallel file systems
such as Lustre and GPFS~\cite{neuwirth2021parallel}. Efforts such as {\tt tf.data}
seek to improve parallel I/O performance by encapsulating the batches as files~\cite{murray2021tf}. 
On the HPC side, studies are trying to integrate AI loaders with I/O libraries having different strategies and performance implications. For instance, Stimulus~\cite{stimulus}, discussed in Section \ref{sec:frameworks}, is a data management library for ingesting scientific data into AI frameworks such as TensorFlow, PyTorch, and Caffe. Other solutions (like HVAC~\cite{9912705}) use caching mechanisms to exploit the node-local storage or near node-local storage to accelerate read I/O and metadata lookups for deep learning applications running on HPC systems.

\subsubsection{Data Transformation} Before data is used, it typically needs to
be pre-processed, which may involve feature scaling, resizing, etc. This is true
both for training and inference. Brewer et al. \cite{brewer2020ibench}
studied the life-cycle performance of distributed inference requests and showed
that pre-processing on the CPU can be the most expensive
component in the inference life-cycle. Recently, there have been efforts to
speed up pre-processing using GPU. For example, Schifferer et al.~\cite{schifferer2020gpu} were able to speed up their feature engineering
pipeline by using a combination of RAPIDS CuDF, Dask, UCX, and XGBoost on a single machine with multiple V100 GPUs.

\subsubsection{Training} Model training is performed either on a single device
or multiple devices in a distributed way, which is generally implemented as
\textit{data-based} parallelism,  \textit{model-based} parallelism, or both.
Data-based parallelism methods shard the training data, such that synchronous
parallel processes each use a portion of the training data, and either
collective communication libraries -- such as NCCL, MPI, or Gloo -- are used
to perform an all-reduce to average the gradients during training \cite{sergeev2018horovod}, or RPC-based methods such as gRPC are used to communicate the gradients. Model-based parallelism may be required when the \textit{memory footprint} of the model exceeds the capacity afforded by the GPU. Additional techniques such as Tensor-parallelism and Pipeline-parallelism extend this approach by partitioning across or within the model's layers. 

The performance of AI models is sensitive to floating point number precision, and there exists
support from software stacks on diverse systems to support various formats
such as half-precision to double-precision. Mixed-precision enabled
implementations vary the precision per layer and can be enabled on the fly
, such as the Automatic Mixed Precision (AMP) on Nvidia GPUs. Benchmarks play a pivotal role in evaluating the performance of training such models, which we will further discuss in Section \ref{sec:benchmarking}. 

\subsubsection{Model Deployment} Model deployment involves optimizing the model for inference performance, which may involve freezing weights, fusing network layers, quantization, and calibration \cite{sze2017efficient}. Tools such as TensorRT for NVIDIA accelerators and OpenVINO for Intel accelerators may perform such optimizations for specific targets. Brewer et al. \cite{brewer2021streaming} show a 10X speedup using TensorRT to deploy a MobileNetV2 model on an NVIDIA T4 GPU. Inference is performed asynchronously parallelly using in-memory approaches such as TensorFlow C API  or distributed inference serving approaches such as TensorFlow Serving or RedisAI \cite{partee2022using}. 

Large-scale inference, such as required for LLMs as Transformers, poses additional challenges. Pope et al. \cite{pope2022efficiently} studied the inference performance of Transformer models with 500B+ parameters and elucidated the challenges involved in efficiently scaling inference for LLMs. First, because the models are generative, they proceed sequentially in time, so subsequent inferences depend on the former ones. Second, models with a large \textit{memory footprint}, exceeding the capacity of a single accelerator, need to be partitioned across multiple accelerators, resulting in additional communication overhead. They study optimal ways of partitioning the models to maximize inference efficiency and demonstrate scaling the Transformer model across 64 TPU v4 chips. 

The speed and efficiency of these inference processes are heavily influenced by how multiple components are coupled (e.g., how the HPC simulation is coupled to an AI model) and whether the inference is executed in memory or sent to a remote inference server on a CPU or GPU. The inference latency, a critical factor in these scenarios, is affected by the network and interconnect performance and the capability of the AI accelerator or GPU. Fast and efficient inference is thus a culmination of optimized model deployment strategies, appropriate use of hardware resources, and effective management of network and interconnect systems. Using benchmarks is crucial to assessing these complex challenges in model deployment and large-scale inference. This aspect, including how benchmarks inform strategies for scaling and optimizing inference performance, will be elaborated in Section \ref{sec:benchmarking}.

\revised{\subsubsection{Simulation} Traditional scientific HPC applications, such as computational fluid dynamics (CFD), or in general finite different, or finite element simulations, require handling the challenges of sustaining performance at scale both at the intra-node (OpenMP) and inter-node (MPI) level, as well as optimizing for on/off loading to the GPU (CUDA/HIP)~\revised{\cite{zhou2020collectives, kondratyuk2021gpu, gonzalez2023heterogeneous}}. Efficient use of hardware also requires handling the challenges of compute-bound versus memory-bound applications. There are performance challenges stemming from inter-process communication involving both collective operations and halo exchanges. On the GPU side, depending on the architecture there may be performance implications whether the network adapters are bound to the GPU directly for direct GPU-GPU communication across nodes (e.g., RDMA) and also based on whether the GPU has to access node RAM or has its own memory.}

\revised{\subsubsection{Data production} Data production involves the generation and storage of large datasets produced by the simulations. On current, large-scale HPC platforms, data moving from compute nodes to storage pass through several layers in a distributed hierarchy of resources (e.g., memory, burst buffers, I/O nodes, filesystem servers). The first critical bottleneck is due to file I/O of such datasets, which may be mitigated, depending on the access pattern, by efficient parallel I/O implementations. At the lower level, most systems have a parallel file system such as Lustre or GPFS, on top of which additional I/O libraries may be exploited, such as ADIOS2, HDF5, and netCDF. Current I/O libraries encounter challenges when staging data between HPC and AI due to their contrasting data access needs. HPC prioritizes high-bandwidth, often write-optimized, I/O for substantial large transfers across parallel file systems. Conversely, AI/ML workloads demand fast, random access to numerous small data samples for training, emphasizing read efficiency. This fundamental conflict in optimization-high-bandwidth sequential (HPC) versus low-latency random (AI)-makes current libraries less efficient in serving both application types optimally during data staging.}

\subsection{Coupling Approaches and Protocols}

Having discussed the performance implications of individual workflow components, we now focus on the broader context of AI-coupled HPC workflows. In particular, we will explore various coupling approaches used in AI-coupled HPC workflows, each utilizing specific protocols and middleware patterns tailored to their use cases. 
To clarify, an \textit{approach} refers to the overall method or strategy used for coupling components, including both the interaction patterns (e.g., data-flow and control flow types and directions) and coupling patterns (e.g., concurrency and dynamism requirements), whereas the \textit{protocol} describes the implementation details of how the data and control messages are exchanged.
For instance, inference deployments often employ request-response type interaction patterns, such as HTTP or gRPC, which facilitate communication between client applications and server-side inference models. For example, SmartSim deployments of \textit{digital replicas} use the the Redis Serialization Protocol (RESP) to perform request-response inference calls.
In contrast, distributed deep learning architectures using in \textit{adaptive execution} typically rely on ring-all reduce type coupling patterns, using protocols like MPI, NCCL/RCCL, or Gloo, to efficiently handle gradient averaging across multiple nodes. More traditional machine learning applications, like those found in data analytics, tend to use map-reduce style interaction patterns, exemplified by frameworks such as Spark or Hadoop, for processing and analyzing large datasets. Additionally, \textit{distributed models} in edge computing or IoT scenarios may adopt publish-subscribe interaction patterns, with protocols like MQTT (used in systems like KubeEdge), to manage data flow between various distributed devices.

To exemplify performance differences between types of distributed training implementations, consider that early versions of TensorFlow, which used the parameter server (request-response) architecture for distributed training, achieved only 50\% parallel scaling efficiency, whereas early versions of Horovod using ring-all reduce achieved 90\% parallel scaling efficiency \cite{sergeev2018horovod}. In addition, implementations such as MPI can take advantage of InfiniBand, whereas protocols such as gRPC only support communication over Ethernet. 

The critical insight to remember is that the performance of a specific workflow is substantially influenced not only by the efficiency of its elements but also by the way those elements are interconnected and integrated. 
Several studies have attempted to investigate how the coupling of workflow components affects overall performance \cite{brewer2021production, yin2022strategies, boyer2022scalable, orland2022case, serhani2024phydll, liu2023nnpred, willard2020integrating}.

Regarding protocol performance, Brewer et al. \cite{brewer2021production} benchmarked inference performance using three different inference serving techniques: HTTP over InfiniBand via a Python-based HTTP server, gRPC over Ethernet via TensorFlow Serving, and the Redis serialization protocol (RESP) via RedisAI. They experimented with several different approaches for load balancing, including using a thread pool, an HAProxy round-robin load balancer, and using MPI. Achieving good scalability required maximizing the batch sizes, which were on the order of 262k samples, and running a single load balancer per node, which round-robinned the inference requests to the multiple GPUs on that particular node (i.e., strong coupling). They demonstrated one million inferences per second for their model on 192 GPUs on the Summit supercomputer. Boyer et al. \cite{boyer2022scalable} extended this approach to benchmark \textit{digital replicas} in online inferencing mode. They noted that na\"ive methods of integrating the simulation with the AI model caused substantial slow-downs, which were significantly mitigated by employing the previously mentioned inference server and load balancing techniques.

\begin{figure}[hbtp]
    \centering
    \includegraphics[width=0.85\textwidth]{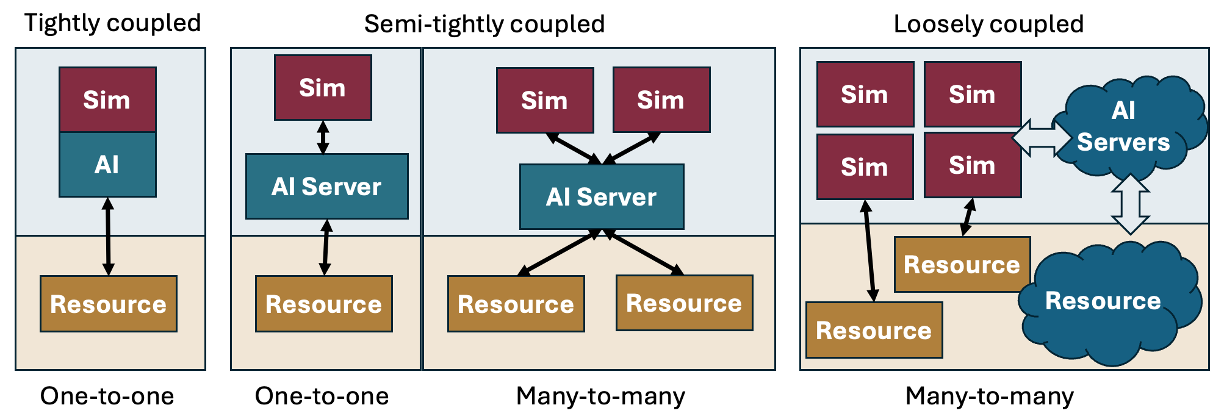}
    \caption{Various strategies for AI-HPC coupling \cite{yin2022strategies}. Tightly coupled systems embed the AI inference in-memory with the simulation. The resource here represents a type of AI accelerator such as GPU or ASIC. The simulation can be decoupled from the AI model via an inference server. Semi-tightly coupled systems may integrate the components within a single compute node or across multiple nodes on a single supercomputer. Loosely-coupled integrations distribute the components across disparate resources, e.g., in the cloud.}
    \label{fig:coupling}
\end{figure}

Regarding coupling strategies, Yin et al. \cite{yin2022strategies} investigated coupling a Monte Carlo simulation to a machine-learned surrogate model using both TensorFlow C++ API and RedisAI inference server. They studied the effect of coupling on inference performance by defining three different coupling types: (1) tightly coupled---in-memory inference with one-to-one mapping between process and accelerator, (2) loosely coupled---distributed inference service on different nodes with many-to-many mapping between the simulation and the model inference server, and (3) semi-tightly coupled---running on same node using inference serving approach, which can have one-to-one, many-to-one, or many-to-many mapping. This is depicted in Fig. \ref{fig:coupling}. They show that tightly coupled deployments are preferred for latency-bound applications, whereas loosely coupled deployments provide better usability via the easy-to-use SmartRedis API. 

\subsection{Motif Performance} 

We now analyze performance issues for each of the motifs introduced in Section~\ref{sec:motifs} regarding their components and coupling methods. Table \ref{table:performance} summarizes the typical performance bottlenecks for each motif and associated implementation. 

\begin{table}[hbtp]
\centering
\begin{tabular}{|l|p{5.1cm}|p{5cm}|}
\hline
\textbf{Execution Motif} & \textbf{Coupling between HPC simulation and AI component} & \textbf{Typical Bottlenecks} \\ \hline
\steering & In-situ coupling, loose or tight coupling strategies & Data transfer between simulation and AI, latency in decision-making \\ \hline
\multistage & Chaining of components, use of intermediate data storage & Data transfer between stages, time spent in each stage \\ \hline
\inverse & Iterative feedback loops, surrogate-assisted optimization & Convergence of optimization algorithm, computationally expensive simulations \\ \hline
\replica & Co-training of models, hybrid modeling approaches & Model integration and communication, training physics-informed AI models \\ \hline
\distributed & MQTT, IoT protocols (e.g., ZigBee), streaming data & Edge-to-HPC network latency, limited memory and compute capacity of edge devices \\ \hline
\adaptive & MPI, NCCL, use of data parallelism or model parallelism & Interconnect performance, network size variability leading to long-job impediment (NAS) \\ \hline
\end{tabular}

\caption{Summary of execution motif characteristics, performance bottlenecks, and coupling strategies.}
\label{table:performance}
\end{table}

\subsubsection{\steeringTitle} 

The \steering motif uses machine-learned models to steer ensembles of simulations. Colmena is an example of such a framework for implementing such a task on HPC and has been demonstrated for molecular design \cite{ward2021colmena}. This type of architecture is similar to distributed reinforcement learning applications, designed to maximize a reward function by running many simulations (environments) in parallel \cite{liang2021rllib}. Dataflow performance is the main challenge for these workflows, especially with a centralized server for task planning that can become a performance bottleneck.

In the \steering motif, data is ingested from parallel simulations, processed, and used to train machine learning models. These models are then evaluated for effectiveness in guiding simulations and deployed in an HPC environment to steer ensembles. For example, Colmena has three processes: a thinker, a task server, and a worker. The thinker communicates with the task server using a request-response architecture, specifically Redis queues, which use the Redis Serialization Protocol (RESP). Workers are scaled up on the system using Parsl \cite{babuji2019parsl}.

\subsubsection{\multistageTitle} 

Campaigns involved in the \multistage motif are common when integrating several methods at different scales and resolutions. For example, the space of possible molecular configurations might be very large, and any single method cannot explore all possibilities with uniform probability. By using different methods with varying levels of cost and accuracy, the space of possible configurations can be traversed intelligently. 

In addition to dataflow across different ``executables'', 
this gives rise to numerous performance issues. For example, the concurrent execution of multiple different stages with different computational costs and resource requirements results in a significant load balancing and workload management challenge. Furthermore, suppose AI models are used in conjunction with or as a substitute for each stage. In that case, many of the problems intrinsic to other motifs arise (e.g., dataflow performance bottlenecks in the \steering motif).

Computational campaigns such as IMPECCABLE~\cite{saadi2020impeccable} use a master/worker-type architecture to manage tasks, whereas MPI is used to scale horizontally across multiple nodes. Inference workloads tend to be I/O bound due to tens of thousands of compressed files that must be distributed across all the worker nodes. 

\subsubsection{\inverseTitle} 
Inverse design is a technique of calculating the causal factors that produced them from a set of observations. Such approaches in various domains include the search for catalysts \cite{freeze2019search}, nanophotonics \cite{wang2022advancing}, and material design \cite{lu2022inverse}.
\revised{The performance challenges in inverse design stem from both the AI model's capabilities and computational constraints, particularly when coupled with high-performance computing (HPC) systems. AI models must accurately represent causal factors from data, but their scalability and ability to handle high data rates, such as those from synchrotron facilities, present significant challenges. Computationally, traditional methods like pseudo-Voigt fitting are limited, and AI-based approaches like BraggNN~\cite{liu2022braggnn} reduce bottlenecks, yet still require substantial resources. Coupling AI with HPC introduces additional issues, including efficient resource allocation across CPUs, GPUs, and accelerators, managing data transfer between nodes, and ensuring parallelism without performance degradation. Addressing these challenges is crucial to maintaining the efficiency and scalability of inverse design workflows in high-data environments.}
 
\subsubsection{\replicaTitle}

The \replica motif describes the integration of simulations with machine learning models, either in cognitive simulations (CogSim) or digital twins. The performance depends primarily on how the model is integrated into the simulation (coupling type) and compute heterogeneity. As mentioned earlier, this falls into two approaches: (1) in-memory or (2) distributed (via inference server). Jia et al. \cite{jia2020pushing} used an in-memory inference approach via the TensorFlow C API to integrate a molecular dynamics (MD) simulation with a machine-learned model. Their model can efficiently scale up to the entire Summit supercomputer, achieving 91 PFLOPS (Peta FLoating-point Operations Per Second) in double precision and 162/275 PFLOPS in mixed-single/half-precision. For this reason, they won the Gordon Bell Prize in 2020. 

Partee et al. \cite{partee2022using} investigated coupling a climate-scale, numerical ocean simulation to a machine-learned oceanic eddy kinetic energy (EKE) surrogate model on a heterogeneous cluster, where the simulation was running on many CPU compute nodes. In contrast, the AI model was being served on one or more GPU nodes using a RedisAI-based inference server on a single GPU node with multiple CPU-based compute nodes sending inference requests to the GPU node (i.e., weak coupling). Their model achieved 1.86 million online inferences per second. 

Boyer et al. \cite{boyer2022scalable} tested both in-memory inference and distributed inference using TensorFlow Serving and RedisAI for three types of physics simulations. They show that using an inference serving approach can outperform in-memory approaches by enabling multi-threading on the inference server and sending concurrent requests to the same GPU, exploiting advantages akin to hyper-threading. 

\subsubsection{\distributedTitle}

The \distributed motif describes workflows where components of the workflow are typically distributed across disparate resources, called the edge-to-cloud continuum, and may be realized as reinforcement learning, distributed deep learning, online learning, transfer or multi-task learning, federated learning, and lifelong or continual learning workflows \cite{rosendo2022distributed}. 

There are a couple of performance issues to consider for this motif: (1) performing inference at the edge (whether this will be performed via single-stream, multi-stream, or offline mode), and (2) continually retraining the model to keep up with data skew and drift. When inferencing at the edge, possible considerations must be taken because edge devices usually have reduced memory and computing capacity compared with HPC. Such implementations are complicated by the additional policy considerations for implementation. Tools such as FuncX/Globus Compute\cite{chard2020funcx} lower the barrier to deployment of such models by exposing endpoints publicly via Globus, where authentication is done in the cloud. 

Suhas et al. \cite{somnath2021building} discuss the challenges of building a federated ecosystem for interfacing experimental facilities with simulations on HPC. They mention that one of the main challenges in this workflow motif is to ``close the latency gap between computation and instruments''. Yin et al. \cite{yin2023toward} study an example of this approach applied to neutron scattering and also note the performance challenges of transferring large files between the HPC and the experimental facilities. 

In addition to the considerations outlined, Dube et al. \cite{dube2021future} emphasize the evolving landscape of HPC in the context of distributed models, underscoring the convergence of HPC with AI and data analytics across diverse environments. This 'Internet of Workflows' highlights the growing need for computational models that efficiently manage data movement and optimize workflows across cloud, edge, and traditional supercomputers. Understanding and adapting to these trends is crucial for enhancing performance and scalability in distributed AI-HPC systems.

\subsubsection{\adaptiveTitle}

We discuss this motif for three types of implementations: Hyperparameter Optimization (HPO), Neural Architecture Search (NAS), and  LLM training. HPO is typically used to optimize a set of training hyperparameters, such as batch size, learning rate, etc.; this is typically performed by training several neural networks in an asynchronous parallel fashion. Optimization methods generally fall into either random search or Bayesian camps, with novel augmentations such as Hyperband designed to speed up the evaluations by adaptive resource allocation and early stopping \cite{li2017hyperband}. Tools such as KerasTuner may be run distributed using a chief-worker strategy to scale such methods across many nodes on HPC \cite{team2022keras}. 

NAS, on the other hand, is primarily concerned with designing optimal neural network architectures. Two popular methods for NAS include evolutionary algorithms (primarily in the form of genetic algorithms) \cite{real2019regularized} and reinforcement learning \cite{zoph2016neural}. Regarding implementing such approaches on HPC, Young et al. \cite{young2015optimizing} present the Multi-node Evolutionary Neural Networks for Deep Learning (MENNDL), which was used to design a Convolutional Neural Network (CNN) on the Titan supercomputer at Oak Ridge National Laboratory. Martinez et al. \cite{martinez2018deep} developed a similar approach called Deep Learning for Evolutionary Optimization (DLEO) and used it to solve a complex compressed sensing problem for reconstructing vibration spectra. Such implementations are generally map-reduce style architecture implementations, which map many training events to multiple accelerators on HPC, reduce the model performance, generate a new set of candidate networks to evaluate, and so on until convergence. Because of such a pattern, researchers have exploited map-reduce frameworks such as Apache Spark to perform such tunings on HPC \cite{meister2020maggy}. Such implementations become more complex when the models are so large each candidate must be distributed across multiple GPUs, as in the case of training LLMs. Such examples of \textit{adaptive training} utilize asynchronous map-reduce style communications organized by the job scheduler. Because different architectural candidates may take longer to evaluate, the typical bottleneck is waiting on a straggling candidate architecture job before proceeding to the next generation in the optimization process. Another performance implication is the cost of the reduction call for each generation.

LLMs are particularly challenging to train primarily because the models are generally much larger than what can fit within the memory of a single GPU, and the models need to be trained on colossal amounts of data. For example, GPT-3 has 175 billion trainable parameters and is trained on 499 billion tokens \cite{liopenai}; it has been estimated that it would take 288 years to train GPT-3 on a single V100 GPU \cite{narayanan2021efficient}. Considerable developments in data and model parallelism have enabled significant speedup by using data-based parallelism methods, such as Horovod, and a combination of pipeline and tensor model parallelism \cite{zheng2022alpa}. Using such a triad strategy of data, tensor, and pipeline model parallelism along with DeepSpeed by Microsoft, Narayan et al. \cite{narayanan2021efficient} have demonstrated the ability to train an LLM with one trillion parameters at 502 PF/s on 3,072 GPUs. Additionally, with the increasing number of non-traditional AI accelerators, training LLMs require different approaches to leverage the hardware. Emani et al. \cite{emani2023comprehensive} provided a comparative evaluation of training LLM models on diverse AI accelerators. Recent studies have also further explored efficient training strategies and performance implications of up to a trillion-parameter LLMs on leadership-class supercomputers \cite{yin2023evaluation, dash2024optimizing}. Achieving such a feat required overcoming GPU memory limitations and minimizing inter-node communication latencies using a combination of pipeline and tensor model parallelism. 

\section{Open challenges and opportunities}\label{sec:open}

\revised{The integration of AI into high-performance computing (HPC) workflows has unlocked new capabilities but also introduced complex challenges that span both computational and methodological dimensions. In this section, we identify four key challenges that must be addressed to advance AI-coupled HPC workflows. First, there is a critical need for co-designing hardware, middleware, and software to efficiently support both AI-\revised{coupled} and traditional HPC workloads, ensuring optimal utilization of computing resources. Second, managing the massive and diverse datasets generated by AI-HPC workflows necessitates the development of scalable and efficient data handling strategies that address I/O bottlenecks, data movement, and real-time processing. Third, the absence of standardized benchmarking tools for evaluating AI-HPC workflows hinders systematic performance comparisons and optimization efforts across different execution motifs. Finally, enabling seamless interoperability, energy-efficient execution, and adaptive resource management remains a significant hurdle, requiring new approaches to dynamic workflow scheduling and cross-platform integration. Addressing these challenges presents opportunities to enhance the scalability, efficiency, and scientific impact of AI-HPC workflows, ultimately enabling more effective simulations, data analysis, and scientific discoveries.}

\subsection{Hardware-Software Nexus}

One of the main challenges for traditional HPC systems in accommodating AI-HPC workflows is engineering a system adept at performing both traditional simulations and modern AI tasks.  This balance hinges heavily on the performance of the internal interconnect, which is highlighted by the vast contrast in performance results between the HPL-MxP \cite{hpl-ai} and the HPCG \cite{heroux2013hpcg} benchmarks. The former primarily measures FLOP performance, while the latter measures GPU-CPU interconnect and GPU memory bandwidth performance. 

Recognizing this challenge, innovators like SambaNova, Groq, GraphCore, Untether AI, Tachyum, NextSilicon, Cerebras, and Fungible are pushing the envelope, with a central focus on optimizing data movement--a crucial step to boosting efficiency in modern AI training and inference. One of the main ways this is achieved is by ``moving the compute element directly adjacent to memory cells'' (untether.ai). Secondly, special-purpose ASIC chips are designed for inference, such as tensor processing units (TPU), intelligent processing units (IPU), data processing units (DPU), language processing units (LPU), tensor cores, etc. Another factor to consider is the exploitation of reduced or mixed-precision training and inference. Using tensor cores at reduced precision offers order-of-magnitude improvements in inference performance \cite{brewer2021production,brewer2021streaming}, while GPU CUDA cores support training performance at double precision. There does not always seem to be a good general understanding of what level of precision is required for such scientific applications; such largely depends on the type of application.

Understanding diverse AI accelerators is critical to leveraging them for AI-HPC workflows. These accelerators exhibit great diversity in their hardware and software stacks, which enables them to run a subset of AI models efficiently. For example, the Cerebras wafer scale engine has massive compute cores. In contrast, the SambaNova SN30 DataScale system has one terabyte of memory per device, which supports models with large memory footprints, and Groq aims to accelerate inference-based workloads. 

The performance of AI models is impacted by choice of implementation, such as C++ or Scikit-learn for classical ML algorithms or frameworks such as Horovod \cite{horovod}, PyTorch DDP \cite{li2020pytorch}, DeepSpeed \cite{deepspeed},  Colossal AI \cite{li2021colossal}, MagmaDNN \cite{nichols2019magmadnn}, TensorRT\cite{jeong2021deep}, and ONNX \cite{onnx}. Each framework provides unique capabilities to run the AI models efficiently across varying deployment scales such as edge, cloud, and HPC scales. When DL models are run at a large scale, their performance is impacted by the collective communication overheads, and there exist various algorithms imbibed into these frameworks, such as \texttt{AllReduce} for DDP and \texttt{Scatter\_Gather} for DeepSpeed. Specialized libraries such as \texttt{NCCL} provide tunable parameters to further optimize collective communications.

With the increasing model complexity and memory footprint, distributed implementation of AI models \cite{jin2016scale}
is an effective way to deploy models at scale.  These approaches include techniques such as data-parallel, model-parallel, tensor-parallel, and pipeline-parallel implementations in currently used frameworks. 
Ben-nun and Hoefler \cite{distributedDL-analysis} present a survey of approaches for parallelizing various DL networks through parallelism in network training and inference. The authors discuss techniques critical to efficiently running distributed models, such as neural architecture search, asynchronous stochastic optimization, and collective communication schemes.
Mayer and Jacobsen \cite{10.1145/3363554} discuss the challenges and techniques for scalable deep learning on distributed infrastructures, including hardware resource scheduling, methods, and training and model data management. 

\subsection{Data and I/O Challenge}

HPC platforms have settled on POSIX-compliant parallel file systems as the primary storage engine and interface. Even for an object-based solution such as DAOS, a POSIX shim layer is still needed to meet the requirement of more traditional modes in scientific applications. However, large, sequential read-and-write is no longer the dominant workload pattern for many data-intensive AI-\revised{coupled HPC} workflows. Instead, small, random read/write is much more prevalent -- which traditional POSIX-based parallel file systems have struggled to keep pace with. The nominal approach is to provision a node-local storage layer, where a small amount of SSD-backed devices are positioned close to the compute blades and generally outside the storage network used by the parallel file system. The challenge is that decoupling from primary storage necessitates data staging, which can be achieved through policy-driven data tiering or workflow support to streamline the process. The current solution space is not mature enough and thus renders this a particular bottleneck to overcome for data-intensive workflow integration. Projects such as NoPFS and SmartRedis discussed in Section \ref{sec:frameworks} are attempts in the right direction. More recent OLCF-6 RFP \cite{OLCF6} calls for an AI-optimized storage system and highlights the challenges and needs in this domain. 

\subsection{Benchmarking}\label{sec:benchmarking}

To effectively evaluate the performance of AI-\revised{coupled } HPC workflows, benchmarking them on diverse use case scenarios/motifs is essential. Such benchmarks could reveal potential performance bottlenecks and, hence, lead to optimized solutions. Several efforts in this direction include MLPerf \cite{mattson2020mlperf}, HPCAI500 \cite{hpcai500}, SAIBench from BenchCouncil \cite{saibench}, and HPL-AI \cite{hpl-ai}. MLPerf \cite{mattson2020mlperf}, developed by the MLCommons consortium of industry, academia, and research laboratories, is a state-of-the-art benchmark suite that includes various machine learning tasks. The MLPerf HPC benchmark \cite{farrell2021mlperf} has been developed to represent scientific machine learning workloads typical of traditional HPC environments. There is also an ongoing effort to include AI-HPC workflow benchmarks in this suite in collaboration with the MLCommons Science group \cite{mlperf-science}.

In the context of \textit{training} benchmarks, MLPerf \cite{mattson2020mlperf} and HPCAI500 \cite{hpcai500} are notable for their comprehensive approach. These benchmarks often utilize metrics such as ``training time to convergence'' and ``time to train to a specified accuracy'', providing a more nuanced view of performance than traditional metrics like ``time per batch'' or ``number of epochs''. Recent trends in benchmarking have shifted towards measuring the time required to train models to a certain level of accuracy, reflecting a more realistic assessment of AI model efficiency in practical scenarios. However, these benchmarks' major limitations are the smaller dataset sizes and the complexity of the deep learning models in scientific applications, which do not always represent realistic scenarios in HPC centers. The HPL-MxP (formerly HPL-AI) benchmark \cite{hpl-ai}, for instance, implements the LINPACK benchmark at a lower precision. Currently, there is a lack of a comprehensive suite of AI-coupled HPC workflow-based benchmarks covering most of the motifs presented in Section~\ref{sec:motifs}. To leverage these benchmarks effectively, it is important to define performance metrics such as end-to-end execution time, AI model throughput, and accuracy.

In addition to training benchmarks, evaluating \textit{inference} performance is equally crucial in AI-coupled HPC workflows. Methods such as MLPerf Inference \cite{reddi2020mlperf} offer a comprehensive framework for this, measuring sequential inference performance across various modes, including single-stream, multi-stream, server, and offline (batch) modes. Inference benchmarks like smiBench \cite{brewer2021production} are particularly useful for studying offline distributed inference performance on HPC systems. Inference \textit{latency} is a key metric representing the time taken for a batch of data to be processed, with results typically reported in seconds or milliseconds. This metric, along with \textit{throughput} measured in Hz or samples per second, provides a detailed view of the efficiency of AI models during the inference phase. The MLPerf Inference benchmark, for instance, adapts its metrics based on the scenario, using 90th-percentile latency for single-stream inferencing and throughput for offline batch inferencing \cite{reddi2020mlperf}.

What do the motifs teach us about benchmarking? Most of the current benchmarking efforts consider an application or model in isolation. There is a gap in incorporating workflows that help measure end-to-end workflows' performance and the crucial components to accelerate the workflow. Since it may be expensive to use the workflows with massive datasets, expensive simulation runs, and complex models, it may be worthwhile to investigate benchmarks that are representative of the actual production runs besides using synthetic data instead of experimental or simulated datasets. 

Understanding the performance intricacies is extremely challenging as we see an increased adoption of workflow tools in scientific machine learning applications. This challenge could be addressed by leveraging existing benchmarking efforts in ML and HPC and working to design and develop new benchmarks accordingly. As this benchmarking process is expensive regarding resources and human effort, it is worthwhile to invest in building benchmark-friendly versions of end-to-end workflow use cases, which mimic realistic workflow execution with minimal overheads in datasets and the scale of runs.

\subsection{Cross-cutting: Interoperability, Energy-Efficiency and Resource Allocation} 

AI-coupled HPC workflows often require a high-level of \textit{interoperability}, typically entailing connecting/interfacing with multiple endpoints for data, AI models, and workflow integration. There are two emerging patterns of choice: (1)~REST API \cite{enders2020cross}
and (2) message queue-based approach \cite{engelmann2022intersect}. 
Each has pros and cons; currently, no widely accepted interfacing mechanisms exist. Further, time-sensitive patterns and long-term campaign patterns, as highlighted by the IRI blueprint report \cite{brown2023integrated}, demonstrate the need for interfacing beyond a single institution's boundary, which brings a host of related challenges such as federated IDs, authentication, authorization, access control, etc., to implement a true end-to-end workflow integration. These challenges are more pronounced in motifs connecting experiments with HPC simulations (e.g., \distributed and \inverse) and in supporting \steering and \multistage motifs, which is only partially addressed by specialized frameworks such as MuMMI or DeepDriveMD, and more general frameworks such as SmartSim (ref. Section \ref{sec:frameworks}).

The importance of \textit{energy efficiency} in high-performance computing (HPC) has intensified due to the hard limits on maximum power capacity. A critical observation in most AI-HPC workflows is their relatively low utilization of peak performance, which may be due to data movement or other I/O activities \cite{brewer2020ibench}. This sub-optimal utilization prompts a vital question: what should be the ultimate goal in this context? The answer might lie in optimizing how one could optimize productivity (Science/Watt) rather than raw performance (FLOP/Watt). This optimization raises a further dilemma: should the focus be on time-to-solution or throughput? One approach could be accelerating processes through high concurrency levels and GPU utilization. Alternatively, engaging GPUs less frequently might be more power-efficient, leading to a trade-off of lower throughput for energy conservation, albeit with longer processing times \cite{brewer2020inference}. Other research areas for improving the energy efficiency of AI-coupled HPC workflows involve dynamic scheduling policies, data locality, ensemble pruning of unnecessary runs, and optimizations of convolution implementations \cite{ejarque2022enabling}.  Benchmarking \revised{such} workflows will be key for understanding how to optimize energy efficiency. 

The traditional batch-based job scheduling approach to \textit{resource allocation} in HPC is at odds with scientific workflows' iterative and interactive nature. These systems, designed for resource efficiency, struggle with the flexibility and on-demand resource allocation required for modern scientific discovery. Specifically, to meet the requirements of AI-coupled HPC workflows outlined in this paper, HPC resource management must evolve to accommodate the changing demands of scientific research. This could involve more adaptive scheduling algorithms, leveraging cloud resources for elasticity, or creating virtualized environments for better resource control. A shift towards an application-centric approach in HPC resource allocation is necessary to support scientific workflows' diverse and evolving patterns. This ensures that computational resources are used efficiently and infrastructural limitations do not hinder discoveries. Hybrid techniques such as LSF-Kubernetes \cite{milroy2022one} and serverless heterogeneous computing \cite{bruel2024predicting} seek to address this issue.

\section{Summary and Discussion}

This paper surveyed the extensive landscape of AI-\revised{coupled} HPC workflows, categorizing them into six execution motifs commonly used in scientific workflows: \steering, \multistage, \inverse, \replica, \distributed, and \adaptive. We highlighted prominent middleware exemplars of each motif and discussed their respective performance implications, revealing the performance challenges due to the complexity of integrating workflows across disparate resources. Moreover, we identify the main challenges of deploying AI-coupled HPC workflows at scale, such as the need for co-designed facilities and further development of middleware and benchmarks. We also identify opportunities for advancement in each area.
Specifically, our survey identified a significant gap in the available workflow benchmarks to assess the performance of the six execution motifs. As such, we acknowledge one of the weaknesses in our survey is related to this dearth of benchmarks, which limited our performance discussions to simpler workflows. To address this issue, we propose a new suite of benchmarks designed to target each execution motif. 

As the landscape of AI-enabled HPC workflows continues to evolve, this survey provides a snapshot of the current state of the art. It serves as a roadmap for developing \textit{next-generation} workflows. 
For example, there are three general categories of how advanced AI systems such as Foundation Models (FM) \cite{bommasani2021opportunities} or generative AI may be used with or as part of AI-coupled HPC workflows:
(1) Workflows to generate the data and train AI systems; (2) Once the advanced AI system is trained, use those as intelligent ways of
steering the simulation; (3) Using advanced AI systems to create the coupling framework between the AI and the simulation.

\revised{These integrated} workflows also offer benefits for pre-training and fine-tuning of advanced AI systems. For example, in the context of the \adaptive motif,
training LLMs at scale on HPC offers potentially more efficient alternatives
at larger scales and lower costs than training in the cloud due to specific
performance optimizations at scale. In general, the deployment of advanced AI systems within \revised{such} workflows will benefit from the robust computational power of HPC systems.

A significant application of generative AI in AI-coupled HPC workflows is
their potential to generate new workflows and associated job scripts based on
specific application needs, user requirements, and the computing
infrastructure. By understanding the objectives and constraints expressed in
natural language, LLMs can propose optimized workflows, tailor job scripts,
and suggest resource allocations that align with the desired outcomes. This
type of optimal workflow generation enhances efficiency and ensures that
computational resources are utilized most effectively.

Workflows are becoming first-class applications for procuring future DOE HPC systems and implementing the Integrated Research Infrastructure (IRI)~\cite{brown2023integrated}. For example, the NERSC-10 RFP technical specifications \cite{nersc10} highlight three classes of workflows to support scientific research campaigns on the future HPC system: high-performance simulations and modeling workflows, high-performance AI workflows, and cross-facility workflows for real-time steering of experiments. These workflows will likely exhibit one or more or a combination of the AI-coupled HPC workflow motifs introduced in Section~\ref{sec:motifs}. Similarly, the technical specifications of OLCF-6 at the Oak Ridge National Laboratory~\cite{OLCF6} also emphasize the fact that scientific campaigns are inherently formulated as workflows that include ``phases of HPC simulation, high-bandwidth communication, interleaved AI training with simulation and inferencing, and data processing''. In both cases, one of the main objectives in explicitly including workflows in the specifications is to move (part of) the management of such workflows from user space to the software ecosystem operated by the facilities.  A closer and hopefully better integration with the platform should address the performance challenges related to \revised{AI-integrated} workflows that we listed in Section~\ref{sec:performance}. 

Moreover, ESnet conducted a meta-analysis of workflow patterns across DOE Office of Science programs in the context of the forthcoming IRI \cite{dart2023esnet}. This report highlights three main science patterns: time-sensitive workflows requiring near-real-time response across facilities for timely decision-making and steering; data integration-intensive workflows that combine and analyze data from multiple sources; and long-term campaign workflows that require sustained access to resources over years or decades to accomplish a well-defined objective. In that context, the integration of AI and HPC in workflows will likely occur in time-sensitive and data integration-intensive workflows and be an important driver in designing the IRI framework. 

The next step in scientific computing performance, scale, and sophistication will require AI-coupled HPC workflows to be a central component of future scientific campaigns. A thorough understanding of \revised{these} workflows and the development of new benchmarks will create a strong foundation for addressing future advancements in the post-exascale era~\cite{workflow-predictions}. 

\section*{Acknowledgments}
This manuscript has been authored by UT-Battelle, LLC, under contract DE-AC05-00OR22725 with the US Department of Energy (DOE). The US government retains and the publisher, by accepting the article for publication, acknowledges that the US government retains a nonexclusive, paid-up, irrevocable, worldwide license to publish or reproduce the published form of this manuscript, or allow others to do so, for US government purposes. DOE will provide public access to these results of federally sponsored research in accordance with the DOE Public Access Plan (https://www.energy.gov/doe-public-access-plan).

This research used resources of the Oak Ridge Leadership Computing Facility at the Oak Ridge National Laboratory, supported by the Office of Science of the U.S. Department of Energy under Contract No. DE-AC05-00OR22725.

\section*{Declaration of generative AI and AI-assisted technologies in the writing process}
During the preparation of this work the authors used GPT-4 in order to improve grammar structures in certain places. After using this tool/service, the authors reviewed and edited the content as needed and take full responsibility for the content of the publication.

\bibliographystyle{unsrt}

\end{document}